\documentclass[english,preprint,floatfix,english,groupedaddress,superscriptaddress]{revtex4-1}
\usepackage[T1]{fontenc}
\usepackage[latin9]{inputenc}
\usepackage{xcolor}
\usepackage{pdfcolmk}
\usepackage{babel}
\usepackage{array}
\usepackage{booktabs}
\usepackage{units}
\usepackage{multirow}
\usepackage{amsmath}
\usepackage{amssymb}
\usepackage{graphicx}
\usepackage{esint}
\PassOptionsToPackage{normalem}{ulem}
\usepackage{ulem}
\usepackage[unicode=true,pdfusetitle,
 bookmarks=true,bookmarksnumbered=false,bookmarksopen=false,
 breaklinks=false,pdfborder={0 0 1},backref=section,colorlinks=false]
 {hyperref}

\makeatletter

\providecommand{\tabularnewline}{\\}
\providecolor{lyxadded}{rgb}{0,0,1}
\providecolor{lyxdeleted}{rgb}{0.67,0,0}
\DeclareRobustCommand{\lyxadded}[3]{{\texorpdfstring{\color{lyxadded}{}}{}#3}}

 \@ifundefined{textcolor}{}
 {%
   \definecolor{BLACK}{gray}{0}
   \definecolor{WHITE}{gray}{1}
   \definecolor{RED}{rgb}{1,0,0}
   \definecolor{GREEN}{rgb}{0,1,0}
   \definecolor{BLUE}{rgb}{0,0,1}
   \definecolor{CYAN}{cmyk}{1,0,0,0}
   \definecolor{MAGENTA}{cmyk}{0,1,0,0}
   \definecolor{YELLOW}{cmyk}{0,0,1,0}
 }

\usepackage{color}
\usepackage{amsthm}
\usepackage{graphics}
\usepackage{times}
\usepackage{bm}
\usepackage{hyperref}
\usepackage{color}
\usepackage{bpchem}
\definecolor{RoyalBlue}{cmyk}{1, 0.80, 0, 0}

\usepackage{tikz}
\usepackage{pgf}

\hypersetup{
colorlinks=true,
breaklinks=true, 
hyperfootnotes=true,
citecolor=RoyalBlue, 
pdftitle = {},
pdfauthor = {Florian H\"ase}
}

\makeatother

\begin{document}

\title{{\Large{}Machine Learning Exciton Dynamics}}

\author{Florian H\"ase}

\author{St\'ephanie Valleau}

\author{Edward Pyzer-Knapp}

\author{Al\'an Aspuru-Guzik}

\email{aspuru@chemistry.harvard.edu }

\address{Department of Chemistry and Chemical Biology, Harvard University,
Cambridge, Massachusetts 02138, USA}

\maketitle

\section*{Abstract}

Obtaining the exciton dynamics of large photosynthetic complexes by
using mixed quantum mechanics/molecular mechanics (QM/MM) is computationally
demanding. We propose a machine learning technique, multi-layer perceptrons,
as a tool to reduce the time required to compute excited state energies.
With this approach we predict time-dependent density functional theory
(TDDFT) excited state energies of bacteriochlorophylls in the Fenna-Matthews-Olson
(FMO) complex. Additionally we compute spectral densities and exciton
populations from the predictions. Different methods to determine multi-layer
perceptron training sets are introduced, leading to several initial
data selections. In addition, we compute spectral densities and exciton
populations. Once multi-layer perceptrons are trained, predicting
excited state energies was found to be significantly faster than the
corresponding QM/MM calculations. We showed that multi-layer perceptrons
can successfully reproduce the energies of QM/MM calculations to a
high degree of accuracy with prediction errors contained within $\unit[0.01]{eV}$
($\unit[0.5]{\%}$). Spectral densities and exciton dynamics are also
in agreement with the TDDFT results. The acceleration and accurate
prediction of dynamics strongly encourage the combination of machine
learning techniques with ab-initio methods.

\section*{Introduction}

Studying the exciton dynamics of large photosynthetic complexes, such
as the Fenna-Matthews-Olson (FMO) and light-harvesting II (LHII) complexes,
has been a topic of much theoretical and experimental interest in
recent years \cite{Adolphs2006,Ishizaki2009_3,Kreisbeck2011,Krueger_Scholes_1998,Mohseni2008,Olbrich2010,Plenio2008,Shim2012,Scholes_curutchet_2007,Sarovar2010,Moix2011,Coker2010,Cao2009,Cao2010}.
The theoretical community has focused on employing and developing
reduced models to understand and describe the dynamics of these complexes
\cite{Seogjoo2007,Mohseni2008,Plenio2008,Rebentrost2009b,Cao2009,Ishizaki2009_3,Ishizaki2011,Sarovar2010,abramavicius2010,Cao2010,Moix2011,Kreisbeck2011,Mazziotti2011,Ritschel2011,Rebentrost2011,Singh_Brumer2011,singh_brumer_2011,Pachon_Brumer_2012,Vlaming2011,Saikin2011,Mohseni2011,Zhu2011,DeVega_2011,RoEiWo09_058301_}.
These models rely on the knowledge of a system Hamiltonian for the
interacting chromophores as well as spectral densities to describe
the coupling of chromophores to their environments (protein, solvent)
\cite{Valleau2012}. Computing a system Hamiltonian or spectral density
is an arduous computational task due to the large number of degrees
of freedom in these complexes. The most detailed approaches used to
obtain these quantities have been mixed quantum mechanics/molecular
mechanics (QM/MM) or semi-classical simulations \cite{Mercer1999,Tao2010}.
In particular, one QM/MM approach which has become popular in recent
years \cite{Olbrich2010,Olbrich2011,Shim2012,Damjanovic2002} consists
in propagating the nuclei in the electronic ground state of the photosynthetic
complex. This approximation ignores the change in electronic structure
due to excitation of the chromophores. Subsequently, for a subset
of time frames, excited state energies for the chromophores are computed
using a quantum method such as time-dependent density functional theory
(TDDFT) \cite{Shim2012}. The energy trajectories are then employed
to extract system-bath correlation functions and finally spectral
densities to use in exciton dynamics. 

The downside of this approach is the large computational cost. Long
molecular dynamics (MD) equilibration times of several tens of nanoseconds
are required \cite{Olbrich2011JCPL,MenucciCPC2014}. The typical computational
scaling of MD codes with the system size $N$ is $\mathcal{O}(N\cdot\log N)$
\cite{Shan2005}. In contrast, TDDFT calculations scale as $\mathcal{O}(N^{2})$
\cite{DRapDFTbook2012}. Very often calculations need to be repeated
for identical chromophores in similar environments to account for
the effect of small variations. For instance in the case of a single-point
mutation typically one would need to rerun the entire set of simulations. 

In this work we propose an alternative route: using multi-layer perceptrons,
a special class of neural networks, to predict the excited state along
a MD trajectory. Such approaches typically scale as $\mathcal{O}(N)$
and were found to perform significantly faster than TDDFT approaches.
As a test system we consider the Fenna-Matthews-Olson (FMO) complex
of \emph{P. Aestuarii.} We use multi-layer perceptrons as fully connected
neural networks to predict the values of the first singlet excited
state for the chromophores. We train the neural networks on the excited
state energies obtained from QM/MM calculations. Several sampling
methods are used to select the training data for the neural networks.
In particular we tested a sampling method based on correlations of
nuclear positions to improve on the spectral density predictions.
Once trained, the neural networks are employed to make excited state
energy predictions. Then one can build a Hamiltonian from the predictions
and compute the exciton dynamics. 

With optimal neural network training and $12$ trained neural networks
per BChl we predicted excited state energies with errors contained
to $\unit[0.01]{eV}$ ($\unit[0.5]{\%}$) from the neural network
ensemble average. Further, with neural networks trained on data based
on correlation sampling we correctly predict the shape of the spectral
density and observe an error which is squared with respect to the
excited state prediction error. This demonstrates the power of machine
learning in chemistry, as has also been found in recent work where
neural networks were employed to extract other chemical properties
\cite{AnatolePhysRevLett2012,AnatoleTDDFT2015,Anatole2015,Rupp-2012,Ed2015}
.

\section*{Methods and Computational Details}

\subsection{Ground state QM/MM simulations\label{sub:Ground-state-QM/MM}}

\begin{figure}
\begin{centering}
\includegraphics[width=0.9\columnwidth]{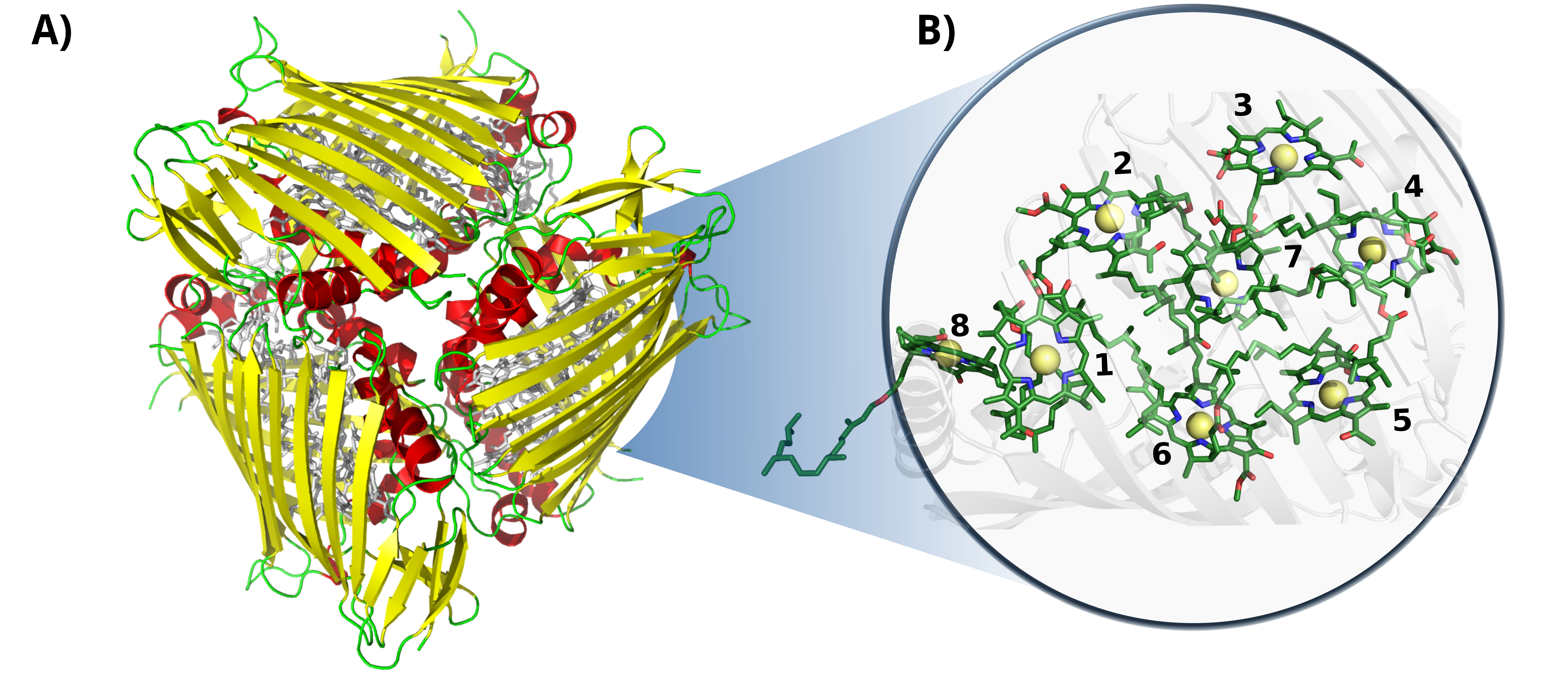}
\par\end{centering}

\protect\caption{Crystal structure of the FMO complex in \emph{P. Aestuarii} (PDB:
3EOJ \cite{Tronrud2009}). (A) 3EOJ trimer crystal structure. (B)
BChls' geometric arrangement in monomer A (residues 360 to 367, corresponding
to sites 1 to 8). Hydrogens are not shown in this representation.
\label{fig:Crystal-structure-of}}
\end{figure}

A semi-classical description of the FMO complex was obtained by combining
ground state MD simulations with TDDFT calculations of the first singlet
excited state, known as the $Q_{y}$ state \cite{Olivucci2005}, at
given molecular conformations along the time-dependent trajectories.
The MD runs were carried out using the NAMD software package \cite{NAMD}
with the AMBER99 force field (ff99SB) \cite{Cornell1995}. The BChl-a
parameters employed are reported in Ref.~\cite{Ceccarelli2003}.
The X-ray crystal structure of the FMO trimer in \emph{P. Aestuarii}
(PDB: 3EOJ \cite{Tronrud2009}, see Fig.~\ref{fig:Crystal-structure-of})
was chosen as initial configuration. The trimer was solvated using
a TIP3P water box \cite{Jorgensen1983}. The minimum distance between
the crystal structure and edges of the solvent box was taken to be
$\unit[15]{\mathring{A}}$ \cite{Olbrich-Box,MenucciCPC2014}. The
charge was neutralized by adding sodium ions. The total number of
atoms in the system was $141624$. The simulation was equilibrated
for $\unit[40]{ns}$ and the production run was $\unit[40]{ps}$ long
with a $\unit[2]{fs}$ time-step. Electrostatic interactions were
calculated with the Particle-Mesh Ewald method. Shake constraints
were used for all bonds containing hydrogen. Simulations were carried
out at $\unit[300]{K}$ and $\unit[1]{bar}$ using a Langevin thermostat
and a Langevin piston as implemented in NAMD.

The time-dependent $Q_{y}$ excited ground state gaps of the BChl-a
molecules were obtained using TDDFT with the PBE0 \cite{PBE0} functional
within the Tamm-Dancoff approximation (TDA) \cite{TammDancoff} using
Q-Chem \cite{QCHEM4}. We employed the 3-21G basis set due to the
high computational cost of these simulations. The $Q_{y}$ excited
state was taken to be the one with the largest oscillator strength
and the orientation of the transition dipole was verified using the
same methodology as in Refs.~\cite{MenucciCPC2014} and \cite{Surya2015}.
Excited state energy distributions are shown in the supplementary
information (see Sec.~\ref{sub:Excited-state-energy}) and trajectories
can be downloaded from \cite{Trajectories2015}. Excited state energies
were computed at every $\unit[4]{fs}$ of the production run. Some
values were excluded based on the oscillator strength / angle criterion
/ failed convergence. The excluded values were at most $\unit[2.15]{\%}$
of the full trajectory.

\subsection{Machine Learning: neural networks training\label{sub:Machine-Learning}}

\subsubsection{Input data representation\label{sub:Input-data-representation}}

Multi-layer perceptrons (neural networks) were used to predict quantum
mechanical excited state energies of BChls in the FMO complex from
the MD classical coordinate trajectory. Neural networks were trained
in a supervised training scheme using the back propagation algorithm
\cite{Rumelhart:1986:LIR:104279.104293}. Excited state energies from
the TDDFT calculations described previously were provided to the neural
network as targets. BChl conformations were represented by Coulomb
matrices as proposed in Ref.~\cite{Rupp-2012}. Both, input and target
feature distributions, were rescaled to a zero mean and a unitary
standard deviation prior to neural network training \cite{Heaton2015}.

By using Coulomb matrices as input features, neural networks can be
trained on a representation of BChls which is translation and rotation
invariant. Coulomb matrices are particularly suitable to describe
BChls in the FMO complex as these molecules do not undergo large conformational
changes within time scales of several tens of picoseconds \cite{Olbrich2011JCPL,MenucciCPC2014,Lilienfeld2015}.
Coulomb matrices were adapted to account for external charges within
and around the represented BChl. The electrostatic influence of particles
in the environment $N$ was described by adding additional Coulomb
potential terms to the corresponding Coulomb matrix entries (see Eq.~\ref{eq:coulombmatrix-1}):

\begin{align}
M_{ij}=\begin{cases}
Z_{i}^{2.4}/2+\sum\limits _{n\in N}\frac{Z_{i}Z_{n}}{|\vec{r}_{i}-\vec{r}_{n}|} & \text{for }i=j,\\
\frac{Z_{i}Z_{j}}{|\vec{r}_{i}-\vec{r}_{j}|}+\sum\limits _{n\in N}\frac{Z_{i}Z_{n}}{|\vec{r}_{i}-\vec{r}_{n}|}+\sum\limits _{n\in N}\frac{Z_{j}Z_{n}}{|\vec{r}_{j}-\vec{r}_{n}|} & \text{for }i\neq j.
\end{cases}\label{eq:coulombmatrix-1}
\end{align}

Partial charges $Z_{i}$ of atoms in the system were taken from the
system topology (Amber 99SB force field \cite{Cornell1995} and Ref.~\cite{Ceccarelli2003}).
Studies have shown that the tails of the BChls have little influence
on the $Q_{y}$ excited state energies \cite{MennucciJCTC2013,MenucciCPC2014}.
Thus, instead of representing the entire BChl in a Coulomb matrix,
the phytyl tail was neglected and only the $90$ atoms closest to
the magnesium in the BChls were represented in Coulomb matrices to
reduce their dimensionality. We included all external partial charges
present in the system to generate Coulomb matrices.

\subsubsection{Neural network architecture, choice of BChl molecule and over-fitting}

We chose to use multi-layer perceptrons (neural networks) with logistic
activation functions and two hidden layers. This set-up has been shown
to perform particularly well for supervised regression problems \cite{RIEDMILLER1994265}.
Optimal neural network hyperparameters were identified from a grid
search. Both the learning rate and the number of neurons in the first
and second hidden layer were varied to find the lowest deviations
between predictions and target data. 

Instead of performing the grid search on each BChl in the FMO complex
individually, only the most representative BChl was used to determine
optimal neural network hyperparameters to reduce the computational
cost. We identified this BChl in terms of shared Coulomb matrix space.
Coulomb matrices of all eight BChls were clustered and compared. We
found that site 3 shares the most Coulomb matrix space with all other
sites (see supplementary information Sec.~\ref{sub:CM-space-cluster}). 

From the grid search, we found that a learning rate of $10^{-4}$
with $204$ neurons in the first hidden layer and $192$ neurons in
the second hidden layer results in the smallest average absolute deviation
of predicted and target excited state energies. 

Target feature over-fitting was avoided by using \textit{early stopping}
\cite{Prechelt1998761}. For all training sessions a total of $4000$
trajectory frames was assigned to the training set as a balance between
information and computational cost. Neural networks were trained on
Intel(R) Xeon(R) CPUs (X5650 @ $\unit[2.67]{GHz}$). Training one
neural network on four cores took about $\unit[(23.9\pm5.0)]{h}$.
Training times for other investigated training set sizes ranging from
$500$ to $5000$ frames are reported in the supplementary information
(see Sec.~\ref{sub:Neural-Network-Benchmarking}).

\subsubsection{Reducing training set redundancies through clustering: taxicab, Frobenius
and ``correlation'' clustering\label{sub:Improving-training-through}}

We employed different methods to select Coulomb matrices and corresponding
excited state energies for neural network training. In a first approach
training set Coulomb matrices were drawn randomly from the entire
trajectory as proposed in Ref.~\cite{Prechelt1998761}. This led
to excited state energy predictions with an average accuracy of $\unit[13]{meV}$. 

However, as BChl conformations in the data set are not uniformly distributed
a training set consisting of randomly drawn Coulomb matrices likely
contains redundant information. MD simulations were carried out in
the NPT ensemble with constant temperature and pressure. Thus, the
BChl conformations are sampled from a Boltzmann distribution \cite{FrenkelSmit2002}.
To avoid selecting many similar conformations and thus similar Coulomb
matrices we performed a cluster analysis on all Coulomb matrices of
the entire trajectory to determine the most distinct Coulomb matrices.
We implemented a Coulomb matrix cluster algorithm following the principles
of the gromos method \cite{Daura-1999}. 

Distances between Coulomb matrices were measured using p-norms. Two
different metrics were applied: $p=1$ (taxicab norm) and $p=2$ (Frobenius
norm). Both clustering approaches resulted in more accurate predictions
of excited state energies with accuracies of $\unit[9]{meV}$ but
the prediction of exciton dynamics remained quite inaccurate (see
Sec. \ref{sub:Exciton-dynamics}).

To improve the prediction of exciton dynamics, we developed a clustering
method based on coordinate correlations in the classical MD trajectory.
We will refer to this approach as the \emph{``correlation''} clustering
method. The $Q_{y}$ state in BChls is mostly distributed along one
of the diagonals which connects opposite nitrogen atoms \cite{MenucciCPC2014}.
Training set frames were thus selected based on high correlations
in the nitrogen root-mean square deviation (RMSD). In particular,
for the $n$-th BChl we sampled from 
\begin{equation}
\left|C_{n}^{\text{RMSD}}(t)\right|^{2}=\left|\left\langle D_{n}^{\text{Nitrogen}}(t)\enskip D_{n}^{\text{Nitrogen}}(0)\right\rangle \right|^{2}.\label{eq:Correlation_function_Bcl-1}
\end{equation}
until 4000 frames with the largest RMSD correlation were selected.
Here, \linebreak{}
$\text{D}_{n}^{\text{Nitrogen}}(t)=\sqrt{\frac{1}{4}\sum_{i=1}^{4}\left\Vert \vec{r}_{n,i}(t)-\vec{r}_{n,i}(0)\right\Vert ^{2}}$
refers to the root-mean square difference in position of the four
nitrogen atoms in the $n$-th BChl at a given time $t$ with respect
to their position in the energy minimized crystal structure at time
$t=0$. This sampling led to a more accurate prediction of the spectral
density (see Sec. \ref{sub:Exciton-dynamics}).

\subsection{Exciton dynamics and spectral densities\label{sub:Exciton-dynamics-and}}

To further compare the predicted excited state energy trajectories
with the TDDFT trajectories we computed the exciton dynamics in the
FMO complex using two different methods. The first is a stochastic
integration of the Schr\"odinger equation as used in Ref.~\cite{Shim2012}.
The second method is the Markov Redfield master equation \cite{Breuer2002}.
Both of the methods are Markovian but the first relies only on the
excited state energy trajectories while the latter also depends on
the spectral density. Here we focus on the sensitivity of these methods
to the changes related to using neural networks rather than to more
subtle questions on dynamics such as those addressed by the comparison
of Markovian with non Markovian methods (e.g. the hierarchy equation
of motion approach \cite{Ishizaki2009_3}).

Finally the spectral density $j(\omega)$, as used in the Redfield
equations of motion, is obtained by normalizing the Fourier transform
of the two-time correlation function as we discussed in Valleau et
al. in Ref.~\cite{Valleau2012}:
\begin{equation}
j(\omega)=\frac{J^{\text{\text{harm}}}(\omega)}{\pi}\quad\text{with}\quad J^{\text{harm}}(\omega)=\frac{\beta\hbar\omega}{2}\int_{-\infty}^{\infty}e^{i\omega t}C^{\text{\text{cl}}}(t).\label{eq:spectral_density-1}
\end{equation}

The superscript ``harm'' refers to the harmonic prefactor which
is needed to connect the QM/MM results to the open quantum system
approach. Here $C^{\text{cl}}(t)$ denotes the classical correlation
function as defined in Ref. \cite{Valleau2012}.

\section*{Results and Discussion}

\subsection{Excited state energy prediction using neural networks\label{sub:Transition-energy-prediction}}

\subsubsection{Acceleration of excited state energy computations with neural networks}

Intel(R) Xeon(R) CPUs (X5650 @ $\unit[2.67]{GHz}$) were used to train
neural networks and predict excited state energies. A total of $12$
neural networks was trained for each of the eight BChls in monomer
A of the FMO complex. Predictions of each of the $12$ neural networks
per BChl were averaged in a neural network ensemble averaging approach
to obtain a more accurate prediction for the excited state energy
trajectory. The spread of predictions of individual neural networks
is given in the supplementary information (see Sec.~\ref{sub:Spread-of-neural}). 

Training sets were generated with all four training set selection
methods (see Sec.~\ref{sub:Improving-training-through}). Predicting
$Q_{y}$ excited state energies for the entire trajectory ($10^{4}$
frames) for one BChl with one neural network took on average $\unit[(3.9\pm0.8)]{s}$
on one core. In contrast, the quantum chemistry calculations using
the TDDFT (PBE0/3-21G) model chemistry required approximately $\unit[60000]{h}$
for the entire trajectory on one core. Required calculation times
to compute excited state energy trajectories for each of the eight
BChls in the FMO complex are reported in Tab.~\ref{tab:Time-required-to-2}. 

\begin{table}
\begin{centering}
\begin{tabular}{|c|ccc|c|c|}
\hline 
\multirow{2}{*}{Method} & \multicolumn{3}{c|}{Training {[}h{]}} & Calculation {[}h{]} & Total {[}h{]}\tabularnewline
 & $t_{\text{Coul}}^{\text{input}}$ & $t_{\text{\text{E}}}^{\text{target}}$ & $t_{\text{train}}$ & $t_{\text{Calc}}$ & $t_{\text{tot}}$\tabularnewline
\hline 
PBE0/3-21G & - & - & - & $480000$ & $480000$\tabularnewline
$\text{NN}_{\text{Corr}}$ & $48$ & $192000$ & $9178$ & $<0.1$ & $201226$\tabularnewline
\hline 
\end{tabular}
\par\end{centering}

\protect\caption{Time required to compute excited state energies ($10000$ frames)
for all eight bacteriochlorophylls (BChls) for TDDFT (PBE0/3-21G)
and neural network (NN) predictions from correlation clustered Coulomb
matrices. Reported times include neural network training ($t_{\text{train}})$
with input ($t_{\text{Coul}}^{\text{input}}$) and target feature
($t_{\text{E}}^{\text{target}})$ generation, excited state calculations/predictions
($t_{\text{Calc}})$ and the total time ($t_{\text{tot}}).$ If trained
neural networks are available, only Coulomb matrices need to be calculated
for neural network predictions, reducing the required time to $\unit[48]{h}$.
Reported times correspond to training a total of $12$ neural networks
independently to obtain ensemble averaged excited state energies.
All reported times refer to calculations on a single core of an Intel(R)
Xeon(R) CPU (X5650 @ $\unit[2.67]{GHz}$).\label{tab:Time-required-to-2}}
\end{table}

If trained neural networks are available, excited state energies of
given BChl conformations can be predicted directly from Coulomb matrices
representing the BChl conformations. Thus, to obtain the excited state
energies of an entire trajectory consisting of $10000$ frames, Coulomb
matrices need to be calculated first. With a calculation time of $\unit[2.19]{s}$
per Coulomb matrix on one Intel(R) Xeon(R) CPU (X5650 @ $\unit[2.67]{GHz})$
core, about $\unit[6]{h}$ are needed to compute all $10000$ Coulomb
matrices. With excited state energy predictions requiring less than
a minute, excited state energy trajectories can be obtained from neural
networks about four orders of magnitude faster compared to TDDFT calculations.

\subsubsection{Accuracy of neural network predictions}

\begin{figure}[h]
\centering{}\includegraphics[width=0.5\columnwidth]{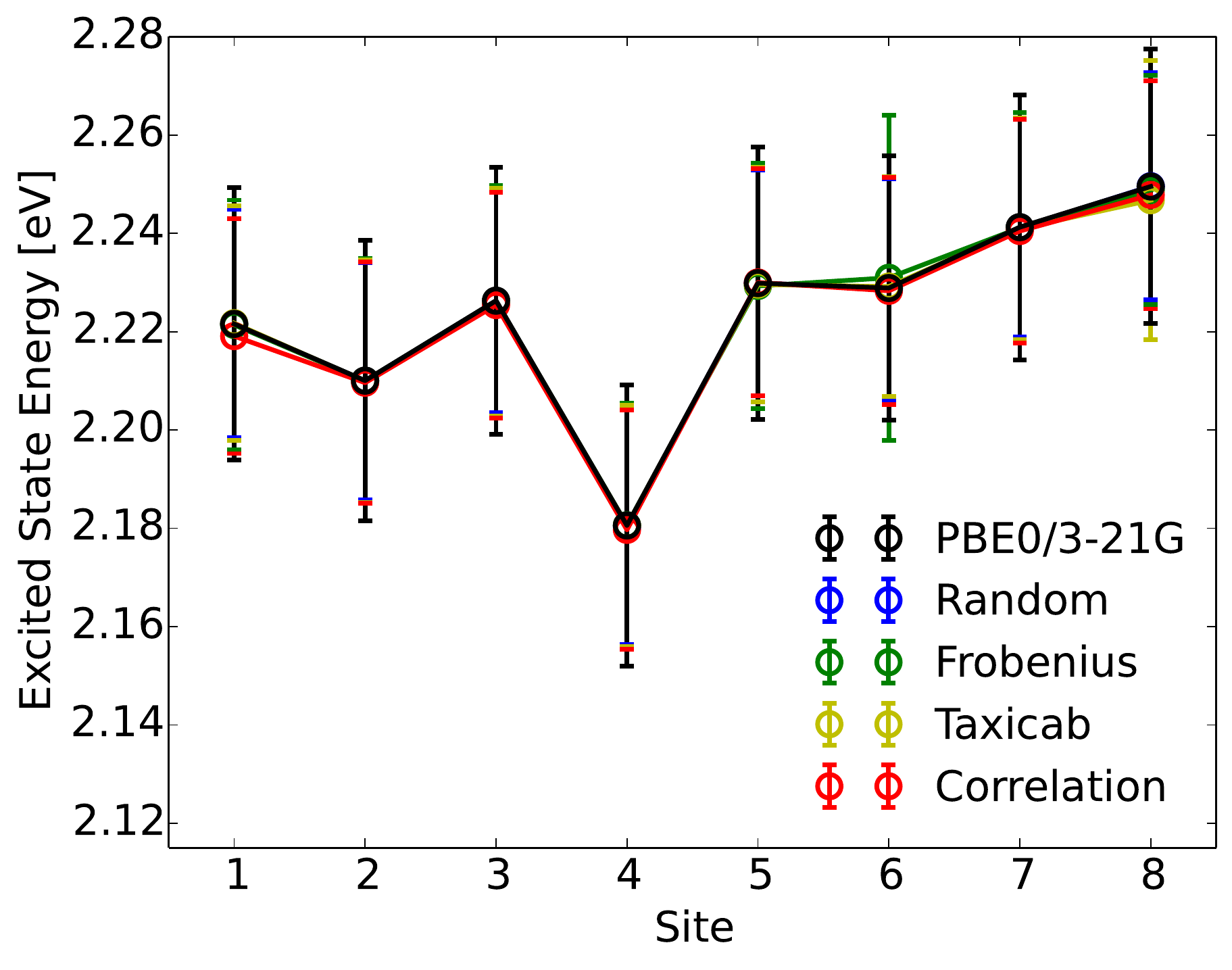}\protect\caption{Mean and standard deviation of $Q_{y}$ excited state energy distributions
for all eight sites obtained from TDDFT calculations (PBE0/3-21G)
and compared to neural network predictions. Neural networks were trained
on Coulomb matrices selected from the classical MD trajectory by the
indicated selection method. Error bars indicate the width of the excited
state energy distribution. \label{fig:Average-excited-state}}
\end{figure}

As can be seen in Fig.~\ref{fig:Average-excited-state} the ensemble
average of $12$ neural network predictions agrees well with the TDDFT
data. Predictions from the average over 12 networks deviate from TDDFT
values by $\sim\unit[0.3]{meV}$ regardless of the site and the input
selection method. When considering excited state energies predicted
from a single network, we observed a deviation of $\sigma_{s,\text{random}}^{\text{single}}<\unit[14]{meV}$
for all sites $s$ when the neural network was trained on randomly
drawn Coulomb matrices. This prediction error can be decreased to
$\sigma_{\text{s,taxicab}}^{\text{single}}<\unit[9]{meV}$ by selecting
the training set based on the Coulomb matrix space clustering taxicab
method (see Sec.~\ref{sub:Improving-training-through}). Frobenius
clustering showed similar deviations. In contrast, predictions from
neural networks trained on correlation clustered Coulomb matrices
show a slightly higher deviation of $\sigma_{\text{s,correlation}}^{\text{single}}<\unit[15]{meV}$
on average.

\subsubsection{Cross-predictions: predicting excited state energies for other bacteriochlorophylls}

\begin{figure*}
\begin{raggedright}
\textbf{A)\qquad{}\qquad{}\qquad{}\qquad{}\qquad{}\qquad{}\qquad{}\qquad{}\qquad{}\qquad{}B)} 
\par\end{raggedright}

\textbf{}%
\begin{minipage}[b]{0.45\columnwidth}%
\includegraphics[width=1\columnwidth]{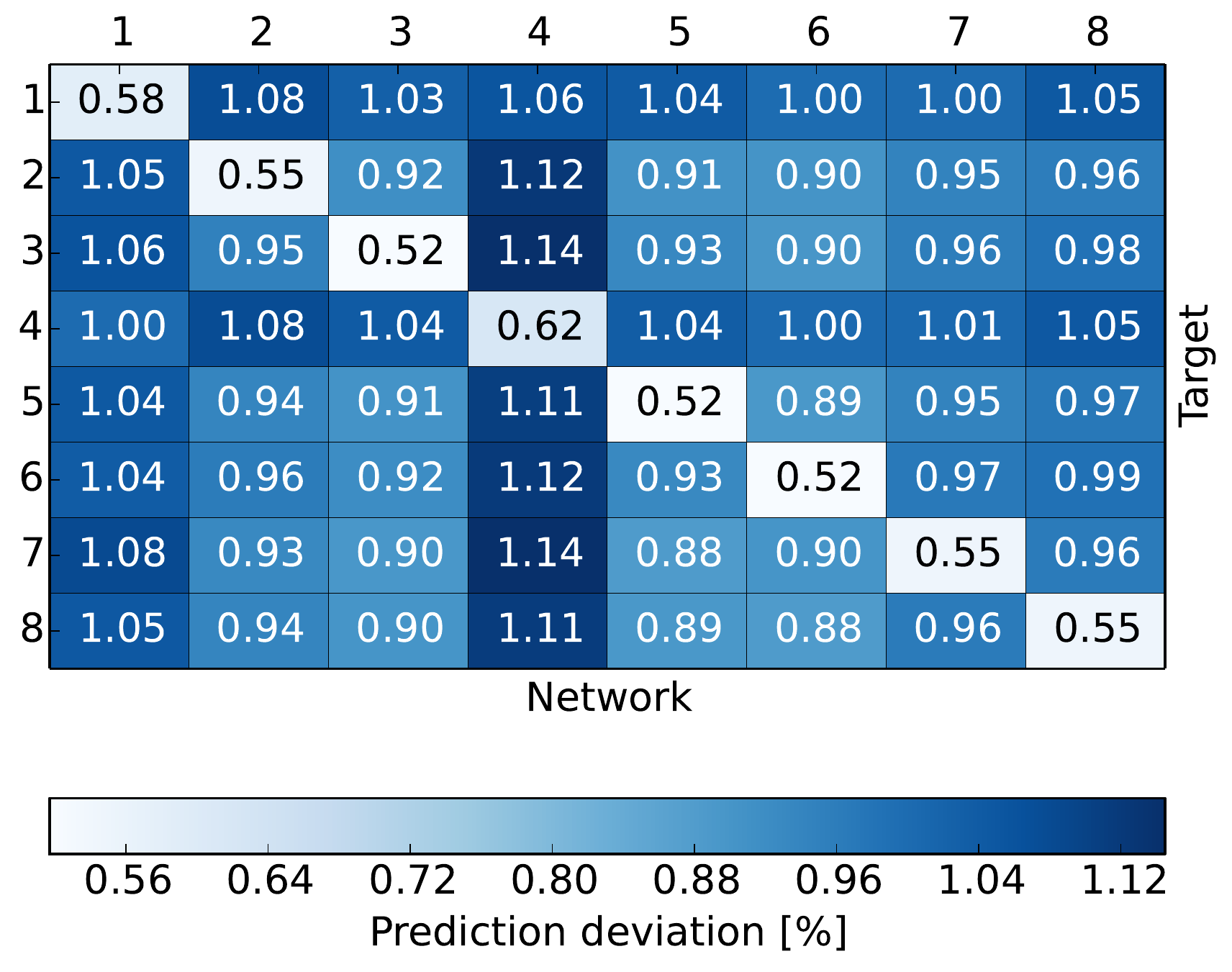}%
\end{minipage} %
\begin{minipage}[b]{0.45\columnwidth}%
\includegraphics[width=1\columnwidth]{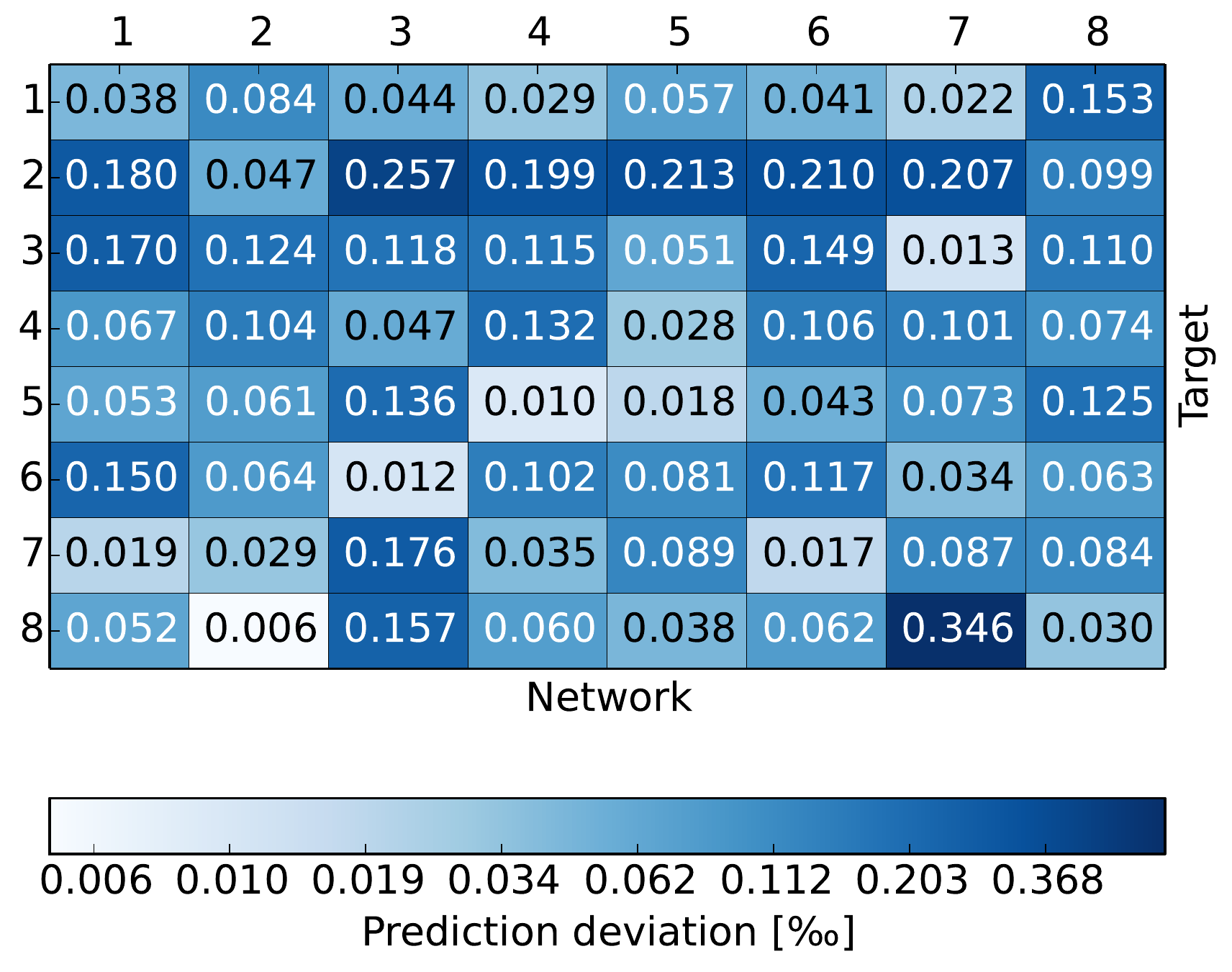}%
\end{minipage}

\protect\caption{Relative absolute deviations of predicted excited state energies from
TDDFT excited state energies. Neural networks trained on one particular
site (indicated by ``Network'') were used to predict excited state
energies of another site (indicated by ``Target''). Panel \textbf{A)}
shows the relative absolute deviation $\bar{\sigma}_{\text{\text{BChl}}}^{\text{rel}}$
of predicted excited state energies from TDDFT excited state energies
for each BChl, $\bar{\sigma}_{\text{BChl}}^{\text{rel}}=\sum_{i}|\epsilon_{\text{BChl}}^{\text{NN}}(t_{i})-\epsilon_{\text{BChl}}^{\text{TDDFT}}(t_{i})|/(N_{\text{frames}}\cdot\langle\epsilon_{\text{BChl}}^{\text{TDDFT}}\rangle)$,
in percent. Panel \textbf{B)} shows the deviation $\sigma_{\text{BChl}}^{\text{mean}}$
of the mean of the predicted excited state energies from the mean
of the TDDFT calculated excited state energies, $\sigma_{\text{mean}}^{\text{BChl}}=|\langle\epsilon_{\text{BChl}}^{\text{NN}}\rangle-\langle\epsilon_{\text{BChl}}^{\text{TDDFT}}\rangle|/\langle\epsilon_{\text{BChl}}^{\text{TDDFT}}\rangle$
in per-thousand. \label{fig:Mean-error-for}}
\end{figure*}

Since all BChl-a molecules in the FMO complex consist of the same
atoms and show similar geometrical conformations, we also used neural
networks trained on one BChl to predict excited state energies of
other BChls in the same monomer. This enabled us to understand how
well the trained network can adapt to changes in the environment from
changes in the Coulomb matrices. We observed that for any clustering
method, the prediction error is about two times larger when performing
this type of cross-prediction, see Fig.~\ref{fig:Mean-error-for}.
Nonetheless, as we see in panel A, the largest observed average absolute
deviation is still below $\unit[1.14]{\%}$ (corresponding to $\unit[25]{meV}$).

\subsubsection{Predicting excited state energies of bacteriochlorophylls in other
monomers}

All neural networks were trained on BChl-a molecules in monomer A.
These neural networks were then used to predict excited state energies
of BChls in the other two FMO monomers. Each neural network predicted
the excited state energies of the BChl corresponding to the one on
which it was trained (i.e. a neural network trained on site 1 in monomer
A predicted site 1 in monomer B and C). 

\begin{figure}[h]
\centering{}\includegraphics[width=0.5\columnwidth]{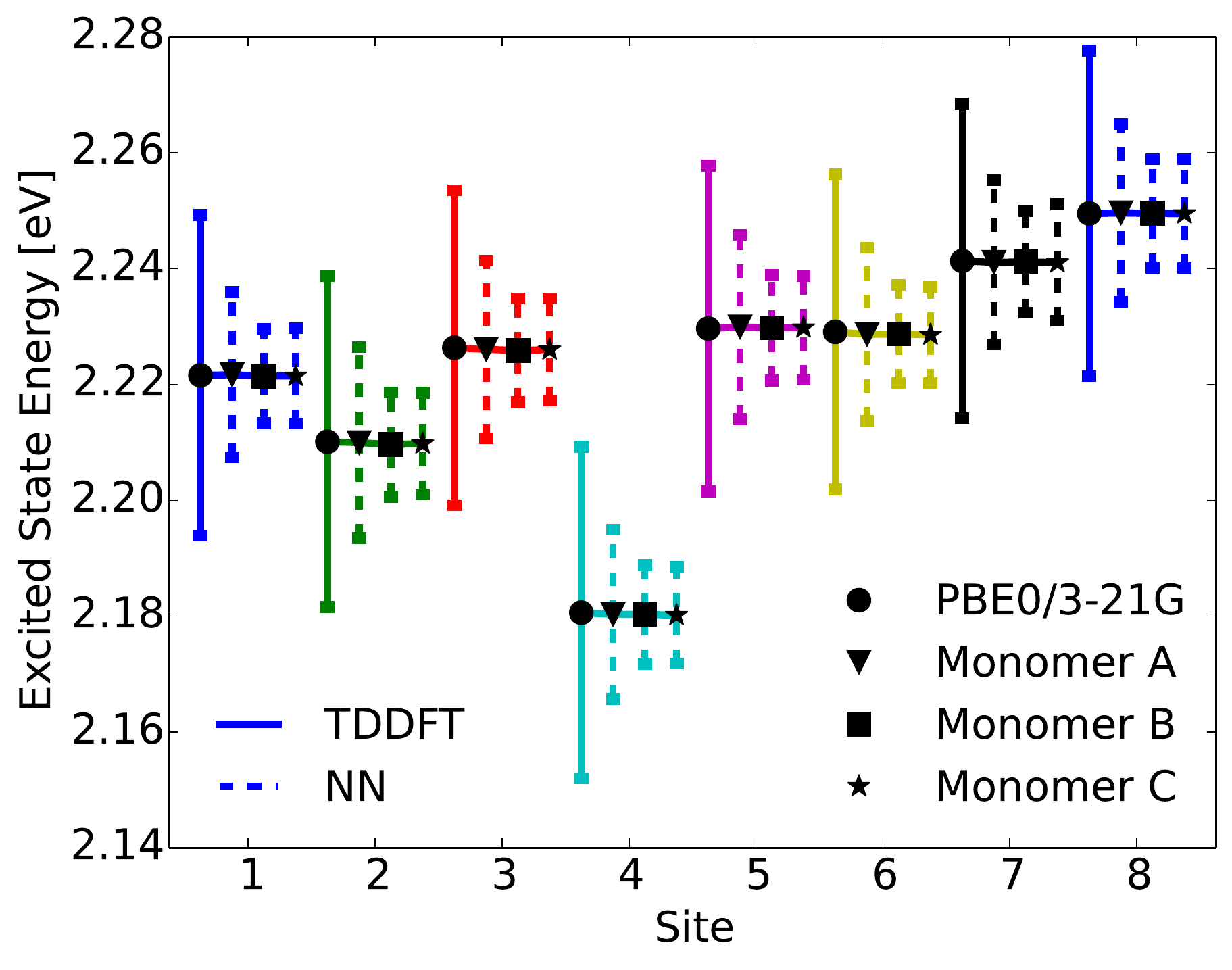}\protect\caption{Mean and standard deviation of excited state energy distributions
for all eight sites obtained using TDDFT PBE0/3-21G (solid line with
circle), for monomer A, and using neural network prediction with training
on 40\% of all Coulomb matrices of monomer A for all three monomers
(dashed lines and triangle, square and star symbol for monomers A,
B, C). Error bars indicate the standard deviation of the obtained
excited state energy distributions. \label{fig:Predicted-average-excited} }
\end{figure}

Due to the fact that the FMO complex is a homo-trimer, similar BChl
conformations should be sampled during the MD simulation in each monomer.
Thus, excited state energy averages are expected to be identical for
corresponding BChls of different monomers, provided that the same
phase space regions were covered in the simulation time. The results
are presented in Fig.~\ref{fig:Predicted-average-excited}. In that
figure, we show the time-averaged excited state energy predicted from
$12$ independently trained neural networks for each monomer as well
as the time-averaged $Q_{y}$ energies obtained from TDDFT for monomer
A. The bars represent the spread of the distribution and not an error.
The predicted distributions are narrower than the TDDFT distributions.
This is probably due to the fact that frames corresponding to energies
at the tails of the distribution are less sampled. Regarding the error
on the other hand, the largest deviation encountered between mean
values of excited state energy distributions is $\unit[3]{meV}$. 

We have found that trained neural networks can accurately predict
TDDFT excited state energies. This allows for a large reduction of
computational time. It is possible to train on a single BChl and predict
excited state energies of other BChls in the same monomer or in different
monomers.

\subsection{Spectral densities and exciton dynamics with neural networks\label{sub:Exciton-dynamics}}

We then used the calculated excited state energy trajectories of all
BChls to obtain information about interactions of the BChls with their
environment by computing spectral densities. In addition, we built
a Hamiltonian to extract the exciton dynamics in the system.

The spectral density $J^{\text{harm}}(\omega)$ (see Eq.~\ref{eq:spectral_density-1})
was computed for all eight sites in the FMO complex from the TDDFT
excited state energies and neural network predicted excited state
energies \cite{Valleau2012}. Spectral densities for each site were
averaged over all BChls to obtain an averaged spectral density $J_{\text{ave}}(\omega)$.
To minimize spurious effects in the Fourier transform, we multiplied
the correlation function by a Gaussian of $\sigma^{2}=0.09\cdot t_{max}^{2}$
with $t_{max}=\unit[1600]{fs}$ as done in Ref. \cite{Valleau2012}.
The Gaussian is normalized to have unitary area in frequency domain
so that in frequency domain this corresponds to a convolution with
a Gaussian with a FWHM of $26\text{cm}^{-1}$.

\begin{figure}[h]
\begin{centering}
\includegraphics[width=0.5\columnwidth]{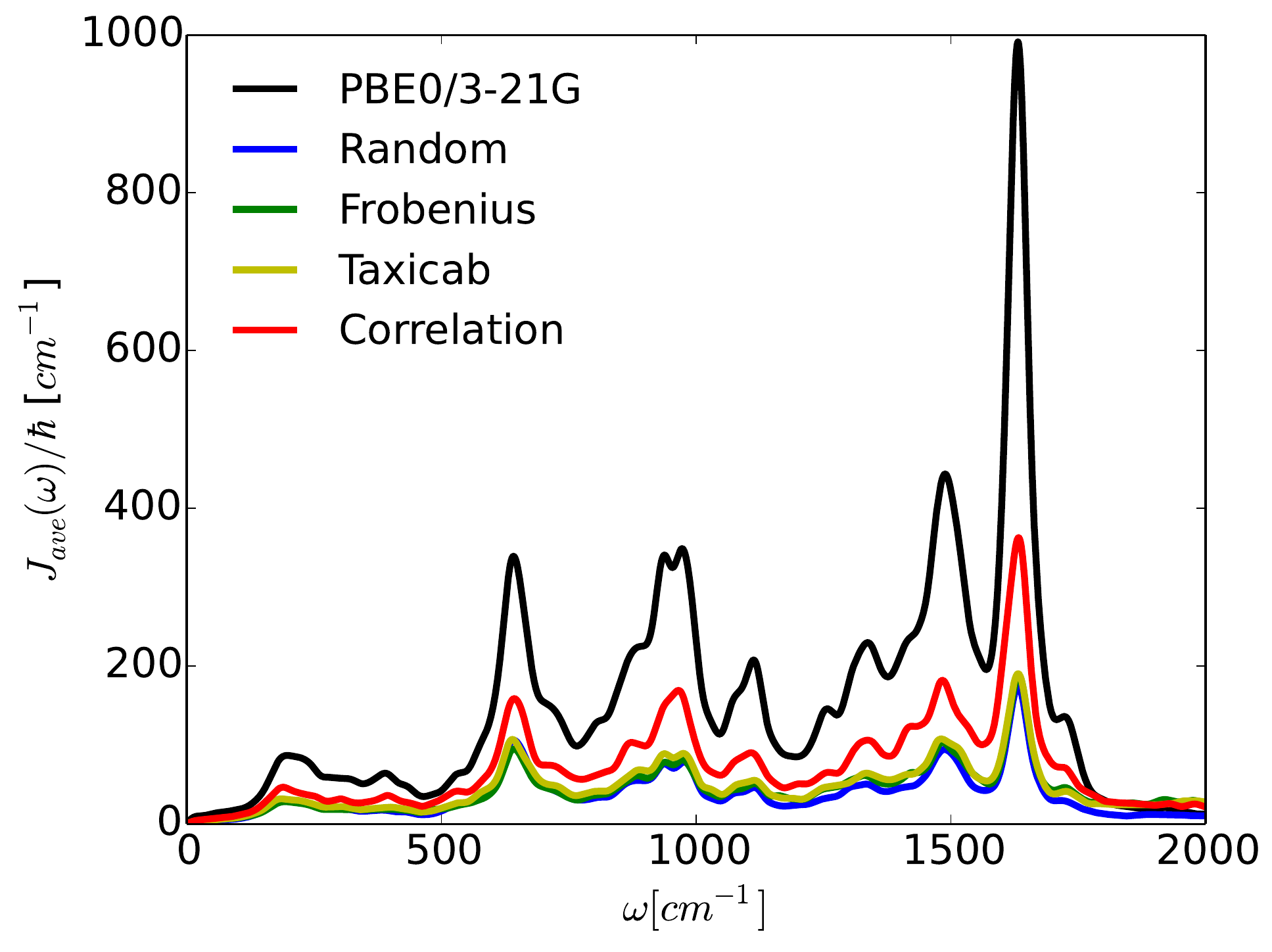}
\par\end{centering}

\protect\caption{Spectral density averages $J_{\text{ave}}(\omega)=\sum_{\text{BChl}}J_{\text{BChl}}^{\text{harm}}(\omega)/N_{\text{BChl}}$.
The spectral densities were computed from excited state energy trajectories
obtained from TDDFT calculations (PBE0/3-21G) and compared to spectral
densities from neural network predicted excited state energy trajectories.
Neural networks were trained on the bacteriochlorophyll they predicted
with the indicated Coulomb matrix selection method. The correlation
method (see Sec.~\ref{sub:Improving-training-through}) gives the
best prediction. \label{fig:Spectral-density-average}}
\end{figure}

In Fig.~\ref{fig:Spectral-density-average} we show the comparison
to our neural network prediction with training and prediction on the
same site and the various Coulomb matrix selection methods. We found
that predicted spectral densities all have a shape which resembles
the overall shape of the TDDFT spectral density. However, the height
and accurate position of the peaks in the spectral density is most
accurately predicted using correlation clustering. 

Spectral densities for each BChl, calculated from TDDFT excited state
energies and excited state energies predicted from neural networks
trained on all input selection methods are plotted in the supplementary
information (see Fig.~\ref{fig:SD_supplementary} in Sec.~\ref{sub:Spectral-Densities}).

Average spectral densities were used to calculate the reorganization
energy \lyxadded{stephanie,,,}{Tue Nov 24 19:14:30 2015}{\linebreak{}
}$\lambda=\int_{0}^{\infty}J_{\text{ave}}(\omega)/\omega\enskip d\omega$.
Comparisons of reorganization energies are reported in Tab.~\ref{tab:Reported-are-the}.
We observe that the smallest deviation between neural network predicted
results and TDDFT results occurs for neural networks trained on correlation
clustered Coulomb matrices. 

\begin{table}
\begin{centering}
\begin{tabular}{ccccc}
\hline 
Method ($\beta)$ & Random & Frobenius & Taxicab & Correlation\tabularnewline
\hline 
$\sigma_{\lambda}\,[\%]$ & $70.0$ & $54.9$ & $53.4$ & $46.6$\tabularnewline
\hline 
\end{tabular}
\par\end{centering}

\protect\caption{We report the percentage deviation of neural network predicted reorganization
energies $\lambda_{\text{NN},\beta}$ from TDDFT (PBE0/3-21G) calculated
reorganization energies $\lambda_{\text{TDDFT}}$ as $\sigma_{\lambda}=|\lambda_{\text{TDDFT}}-\lambda_{\text{NN},\beta}|/\lambda_{\text{TDDFT}}$,
where $\beta$ indicates the method used for Coulomb matrix selections.
Results are given in percent $\%$. \label{tab:Reported-are-the}}
\end{table}

\begin{figure}[h]
\centering{}\includegraphics[width=0.5\columnwidth]{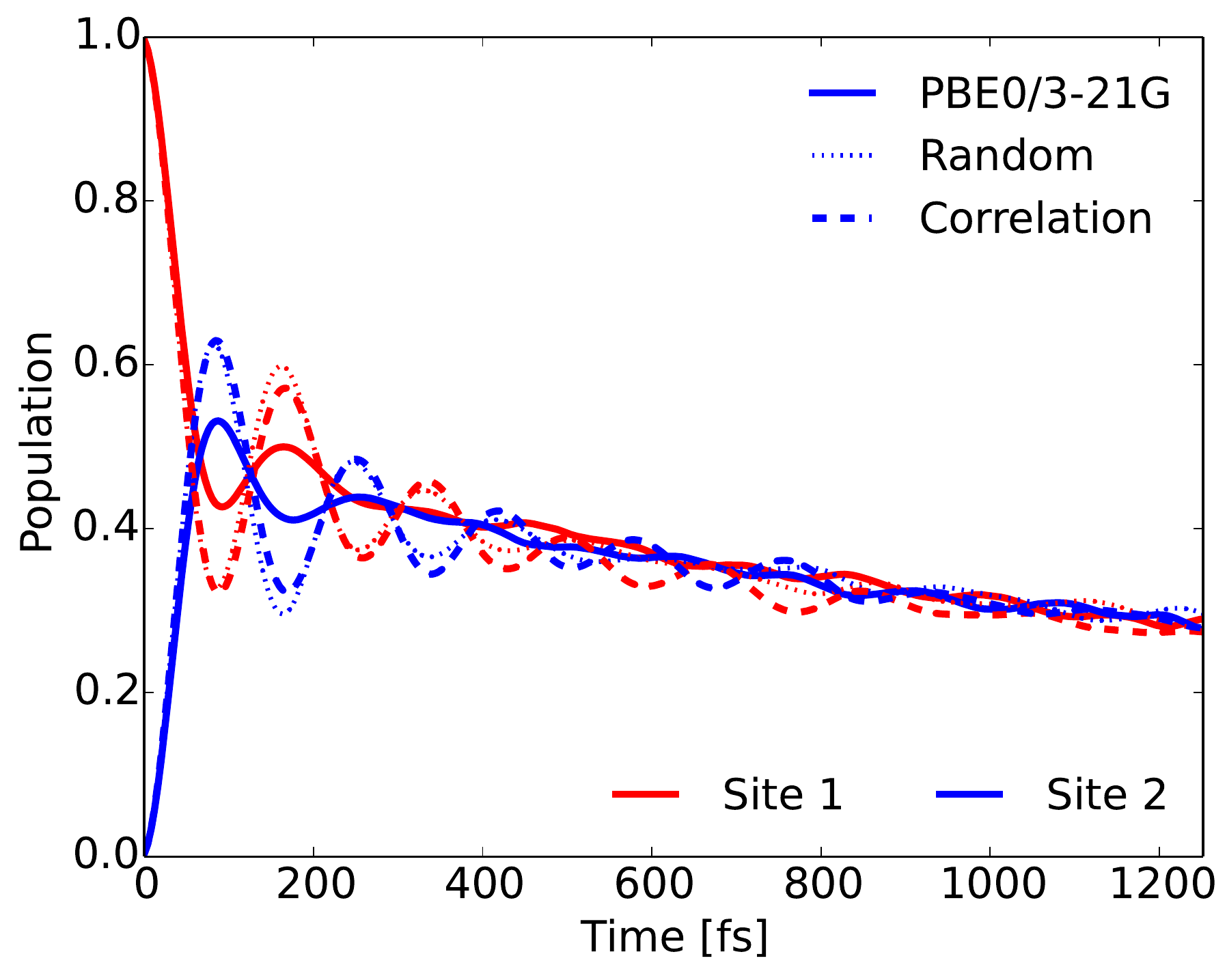}\protect\caption{Population dynamics of the FMO complex calculated at $\unit[300]{K}$
with initial state in site 1. Here only two site populations are shown
but dynamics was carried out for the 8 BChls. Excited state energy
trajectories were obtained from TDDFT calculations (PBE0/3-21G) as
well as neural networks trained on randomly drawn and correlation
clustered Coulomb matrices. \label{fig:Time-evolution-of}}
\end{figure}

We also computed the population dynamics in the FMO complex monomer
with a stochastic integration method \cite{Shim2012}. We averaged
$4000$ stochastic trajectories to obtain converged population dynamics.
The initial excited site was chosen to be site 1 and the dynamics
was propagated for all eight coupled sites at $\unit[300]{K}.$ The
couplings of the Hamiltonian were taken from Ref.~\cite{Surya2015,Aghtar2013}.
Results are shown in Fig.~\ref{fig:Time-evolution-of}. We see that
neural network predictions from neural networks trained on randomly
drawn and correlation clustered Coulomb matrices predict the exciton
dynamics in agreement with TDDFT calculations. 

We also employed the Markovian Redfield method to compute the exciton
dynamics in the FMO complex (see Sec.~\ref{sub:Exciton-dynamics-and})
\cite{Breuer2002}. In this case there is an explicit dependence of
the exciton dynamics on the spectral density. The energies in the
Hamiltonian were taken to be the averages from the TDDFT or neural
network predicted excited state energy trajectories. The same couplings
as for the stochastic integration method were used. 

\begin{figure*}
\begin{centering}
\textbf{A)}\includegraphics[width=0.45\columnwidth]{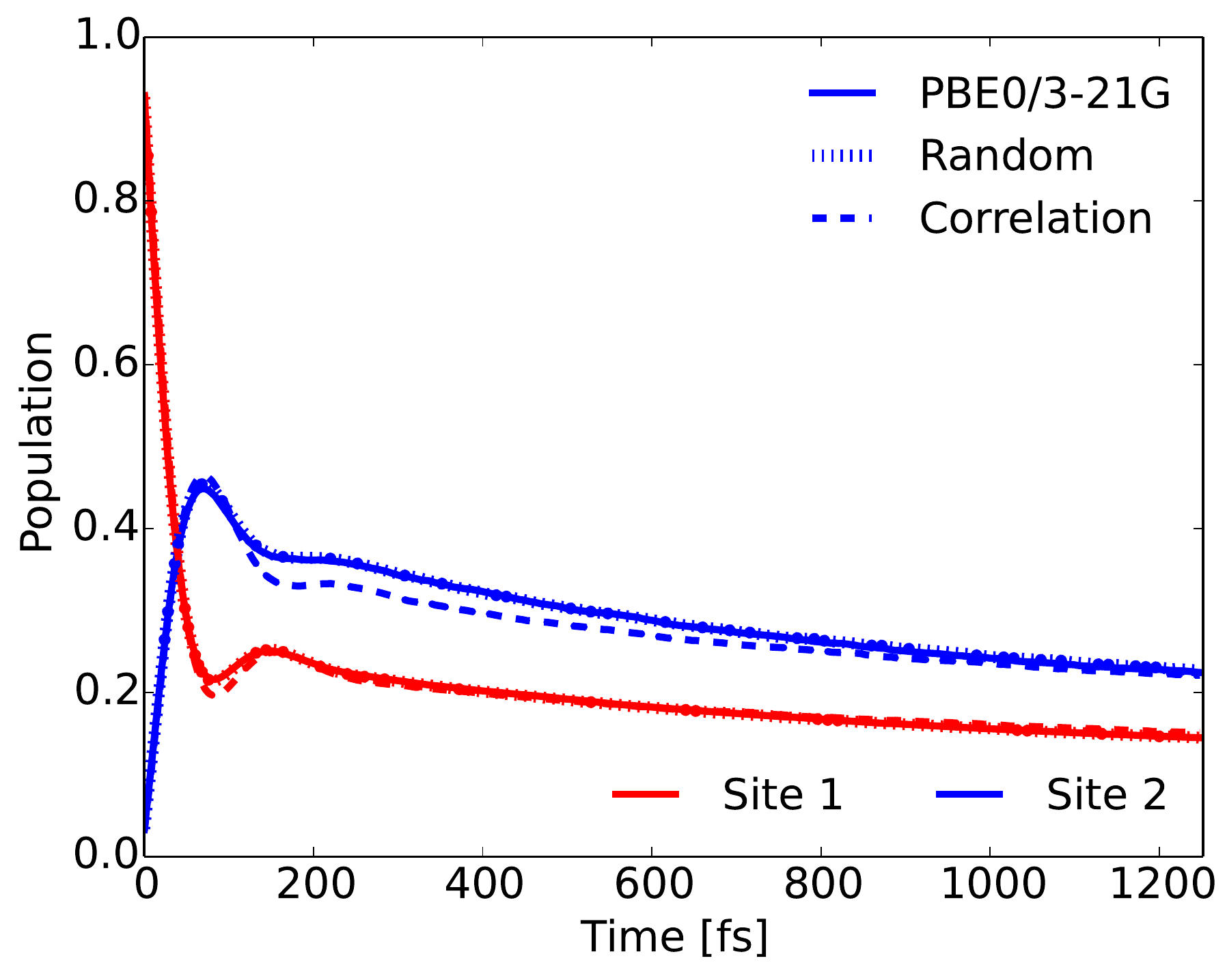}\textbf{B)}\includegraphics[width=0.45\columnwidth]{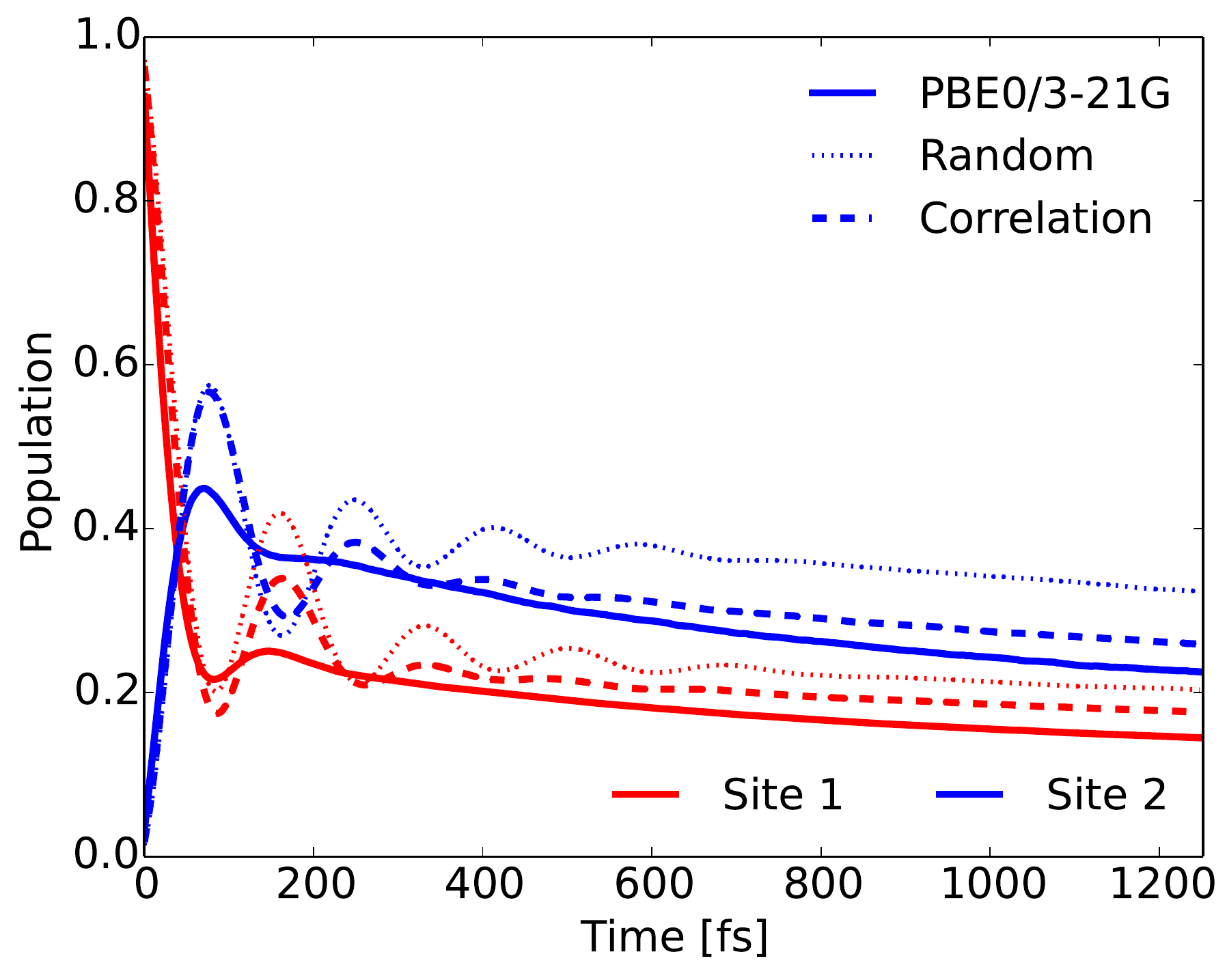}
\par\end{centering}

\protect\caption{Time evolution of the exciton population for BChl 1 (red) and BChl
2 (blue) in the FMO complex calculated from excited state energy trajectories
and average spectral densities using the Redfield method. The initial
state is site 1 excited. Panel A) shows the exciton dynamics for neural
network predicted excited state energies using the same TDDFT calculated
average spectral density in all cases. Panel B) shows the exciton
dynamics with both, excited state energy trajectories and harmonic
average spectral densities predicted by neural networks trained with
the indicated selection method. \label{fig:Time-evolution-of-1}}
\end{figure*}

To investigate the importance of excited state energies and spectral
densities on the Redfield exciton dynamics we calculated the exciton
dynamics in two different ways. First, we computed the exciton dynamics
with neural network predicted excited state energies and the average
spectral density obtained from TDDFT calculations and then we used
the neural network predicted energies as well as the neural network
predicted spectral densities. Results are presented in Fig.~\ref{fig:Time-evolution-of-1}
panel A, for predicted excited state energies and TDDFT spectral densities
and panel B, for excited state energies and spectral densities predicted
by neural networks. We initialized the dynamics with the excitation
in BChl 1 and propagated the dynamics for all eight coupled sites
at $\unit[300]{K}.$

In panel A we see that given a constant spectral density the error
on energies is small and does not strongly influence the exciton dynamics.
The main role is played by the spectral density as can be seen in
panel B. Here we see much larger differences depending on the neural
network sampling method and we notice that the prediction of neural
networks trained on correlation clustered Coulomb matrices agrees
better with the TDDFT exciton dynamics. Deviations of neural network
predicted exciton dynamics from TDDFT calculated exciton dynamics
are listed in detail in the supplementary information (see Sec.\ref{sub:Spectral-Densities}).
Further methods should be investigated to improve the prediction of
the spectral density, the error on the energy should be reduced further
by an order of magnitude to predict spectral densities with the optimal
reorganization energy.

\section*{Conclusion}

The computational study presented in this article showed that multi-layer
perceptrons (neural networks) can be used to successfully predict
excited state energies of BChls in the FMO complex. Using different
methods to select a training data set for neural network training
we were able to generate neural networks which can predict TDDFT excited
state energies with high accuracy ($\unit[0.01]{eV}$). Furthermore,
the neural networks can predict properties such as QM/MM derived spectral
densities or exciton population dynamics. The prediction of excited
state energies using the neural networks is about seven orders of
magnitude faster than TDDFT. If we include training feature generations
we still observe a speed-up of about four orders of magnitude. Even
if neural networks need to be trained first, excited state energies
are obtained in less than half the time needed for TDDFT. 

Based on the observations we made on the FMO complex we recommend
the following procedure to apply machine learning to predict excited
state properties of other systems:
\begin{enumerate}
\item Represent all molecules of interest and their chemical environment
with Coulomb matrices. 
\item Obtain optimal neural network architectures from hyperparameter grid
searches. In particular, it is important identify the most representative
molecule in terms of the space which contains the features used for
neural network predictions.
\item Determine an optimal training set. Training sets for neural networks
with optimal network architecture can be generated by selecting Coulomb
matrices based on properties which are related to the desired quantum
mechanical properties. We observed that selecting Coulomb matrices
which represent Coulomb matrix space clusters improves excited state
energy predictions. However, these do not necessarily work well for
dynamics. 
\item Make predictions. To predict spectral densities and exciton dynamics
with high accuracy, Coulomb matrices should be selected for the training
set if they reveal high excited state energy correlations. We found
that nitrogen RMSD correlations in the BChls are a good indicator
for excited state energy correlations. Thus, we selected Coulomb matrices
based on high nitrogen RMSD correlations. Of course this might be
complicated for some molecules with very delocalized excited states. 
\end{enumerate}
In conclusion, this approach provides a gigantic speedup to ground
state QM/MM. From a neural network trained on a single BChl molecule,
we can predict the excited states of $23$ other molecules in the
system at very low additional computational costs. This will be particularly
useful to speed up or simply to enable the simulation of larger, more
complex and challenging light-harvesting systems. Further, it will
be helpful to study, for instance, the role of small changes in the
environment of the exciton dynamics. Future questions we would like
to address include the possibility of extending the prediction to
other temperatures using for example multi-target machine learning. 
\begin{acknowledgments}
S.V. and A.A.-G. acknowledge support from the Center for Excitonics
and Energy Frontier Research Center funded by the U.S. Department
of Energy under award DE-SC0001088. Computations were run on the Harvard
University\textquoteright s Odyssey cluster, supported by the Research
Computing Group of the FAS Division of Science.
\end{acknowledgments}

\bibliographystyle{pnas2009}
\phantomsection\addcontentsline{toc}{section}{\refname}\bibliography{bibliography_clean}

\clearpage{}

\newpage{}

\section*{Supplementary Information}

\subsection{Coulomb matrix space cluster analysis: choosing the best grid-search
BChl molecule\label{sub:CM-space-cluster}}

The neural networks were trained on Coulomb matrices representing
BChl conformations generated during the classical MD simulation in
a supervised training-scheme. TDDFT excited state energies were used
as training target. To design an optimal neural network architecture
with a minimal deviation between predictions and targets, a grid search
on several neural network hyperparameters was performed. Learning
rate and number of neurons in the first and second hidden layer were
changed step-wise, as reported in Sec.~\ref{sub:Neural-Network-Benchmarking}. 

\begin{figure}
\begin{centering}
\includegraphics[width=0.5\columnwidth]{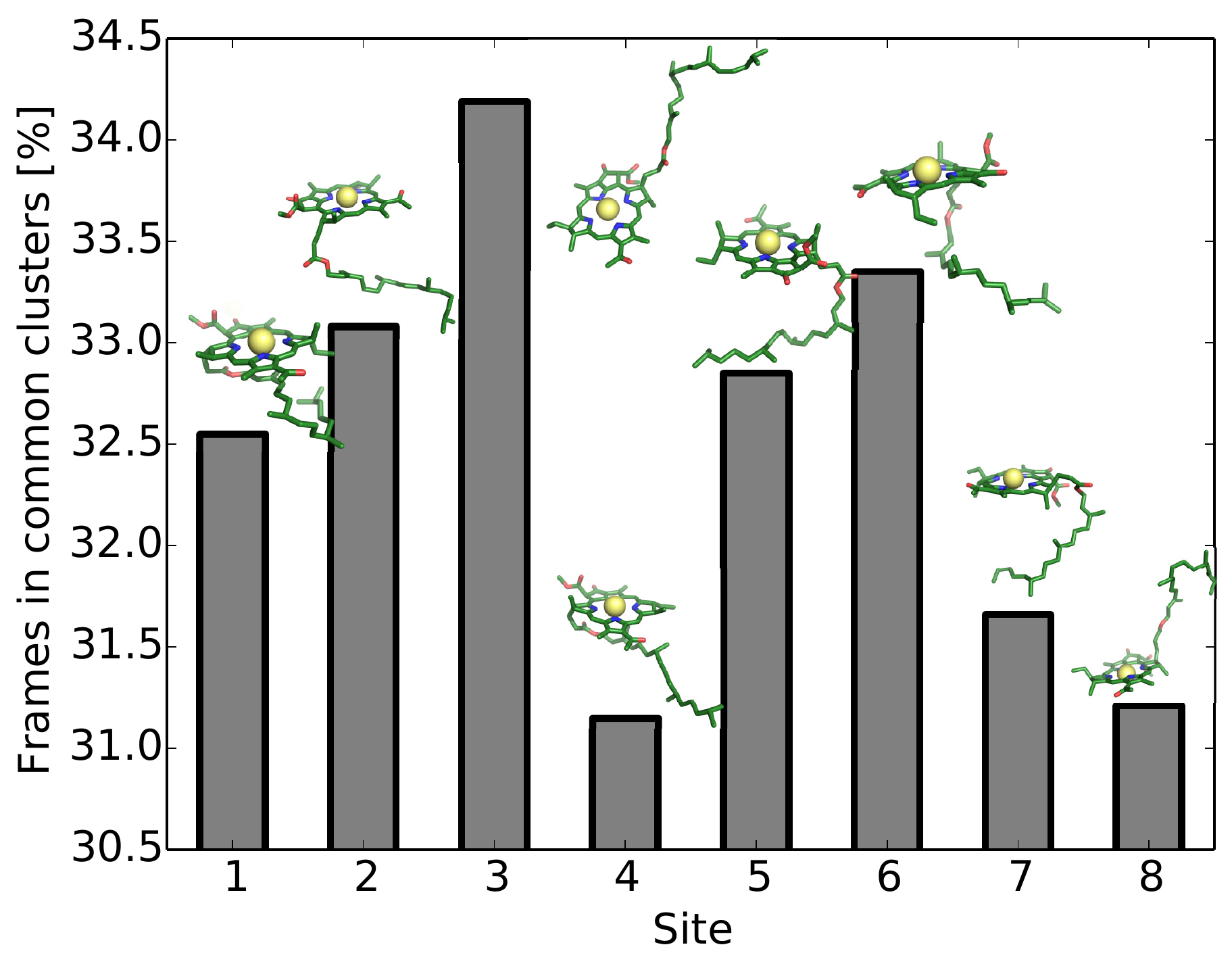}
\par\end{centering}

\protect\caption{Common Coulomb matrix space regions of bacteriochlorophylls observed
during the $\unit[40]{ps}$ production run. We report the number of
frames (in $\%$) of one site contributing to Coulomb matrix space
clusters shared with at least one other site (cluster cut-off: $\unit[90]{e^{2}/\mathring{A}}$).
Stick representations of corresponding representative geometries of
each individual site are shown for comparison. \label{sup_fig:cmSpaceEstimation} }
\end{figure}

We found that BChls in the FMO complex show a high Coulomb matrix
space overlap throughout the entire MD trajectory (see Fig.~\ref{sup_fig:cmSpaceEstimation}).
By determining the site with the highest Coulomb matrix space overlap
with all other sites we were therefore able to limit the number of
grid searches for optimal neural network hyperparameters to one. Optimal
neural network hyperparameters obtained for the most representative
site with the highest Coulomb matrix space overlap were used for all
other sites. 

To identify the most representative site we first started a cluster
analysis on all Coulomb matrices representing site 1 based on the
gromos method \cite{Daura-1999}. Coulomb matrix distances were measured
with the Frobenius norm (see Sec.~\ref{sub:Improving-training-through}).
The clustering cut-off was chosen to be $\unit[90]{e^{2}/\mathring{A}}$
to clearly distinguish Coulomb matrices. Then, for all other sites,
distances of all Coulomb matrices of all frames in the MD trajectory
to all Coulomb matrices representing on particular cluster were calculated.
If the distance was below the cut-off, the Coulomb matrix was attributed
to the cluster. The pool of remaining Coulomb matrices was again clustered
with the same method as site 1. 

Coulomb matrix space overlap of one site with all other sites was
then estimated by counting the number of Coulomb matrices of the considered
site and of Coulomb matrices of all other sites in every individual
cluster. Whenever two sites $i$ and $j$ contributed to the same
cluster different numbers of Coulomb matrices $n_{i}$ and $n_{j}$,
the smaller number of Coulomb matrices $\text{min}(n_{i,}n_{j})$
was added to the Coulomb matrix space overlap estimation of both sites.
This summation was carried out over all sites and all clusters. Quantitative
results are presented in Fig.~\ref{sup_fig:cmSpaceEstimation}.

We observed that of all BChls in the FMO complex, site 3 has the most
shared Coulomb matrix space with all other sites. However, the difference
of shared Coulomb matrix space volumes is small. The highest observed
value ($\unit[34.19]{\%}$ for site 3) is only about $\unit[10]{\%}$
greater than the smallest observed value ($\unit[31.15]{\%}$ for
site 4) and for every site about one third of all Coulomb matrices
lies in shared Coulomb matrix space regions. Hence, we expect neural
network architectures optimized on one site to also work well for
all other sites. This assumption was confirmed by the similarity in
prediction accuracy of neural networks trained on different sites
(see Sec.\ref{sub:Transition-energy-prediction}).

\subsection{Neural network parameter grid search: finding the best network architecture\label{sub:Neural-Network-Benchmarking}}

The prediction accuracy of a neural network is highly influenced by
its architecture. Introducing hidden layers to the neural network
architecture allows for the distinction of data which is not linearly
separable. In this study we used multi-layer perceptrons (i.e. fully
connected neural networks with at least one hidden layer) with two
hidden layers and logistic activation functions. Several neural networks
with different learning rates and numbers of neurons in the first
and second hidden layer were designed and trained on Coulomb matrices
and corresponding excited state energies. We used the back-propagation
algorithm and a supervised training scheme with excited state energies
as target. Overfitting was avoided with the early stopping method
(see Sec.~\ref{sub:Machine-Learning}). Thus we determined an optimal
set of hyperparameters to identify a neural network architecture suitable
for accurate excited state energy predictions for BChls in the FMO
complex from classical MD simulations. 

The entire grid search was performed on BChl 3, which was shown to
represent the most of the Coulomb matrix space covered by all BChl
molecules in the FMO complex during the $\unit[40]{ps}$ production
run (see Sec.~\ref{sub:CM-space-cluster}). Since optimal neural
network architectures were not known beforehand we chose particular
initial hyperparameter values prior to the grid search. Unless specified
otherwise neural networks were trained with a learning rate of $10^{-3}$
and $180$ neurons in the first and second hidden layer. Neural networks
were trained on $3000$ frames to keep the training times during the
grid search reasonably small.

\subsubsection*{Learning rate }

To determine an optimal learning rate we set up five different neural
networks with learning rates of: $10^{-3}$, $5\cdot10^{-4}$, $10^{-4}$,
$5\cdot10^{-4}$ and $10^{-5}$. All of these neural networks were
trained on $\unit[30]{\%}$ of all trajectory frames representing
BChl 3 by Coulomb matrices. The neural networks were designed with
two hidden layers consisting of 180 respectively.

\begin{figure}[!ht]
\centering{}\includegraphics[width=0.5\columnwidth]{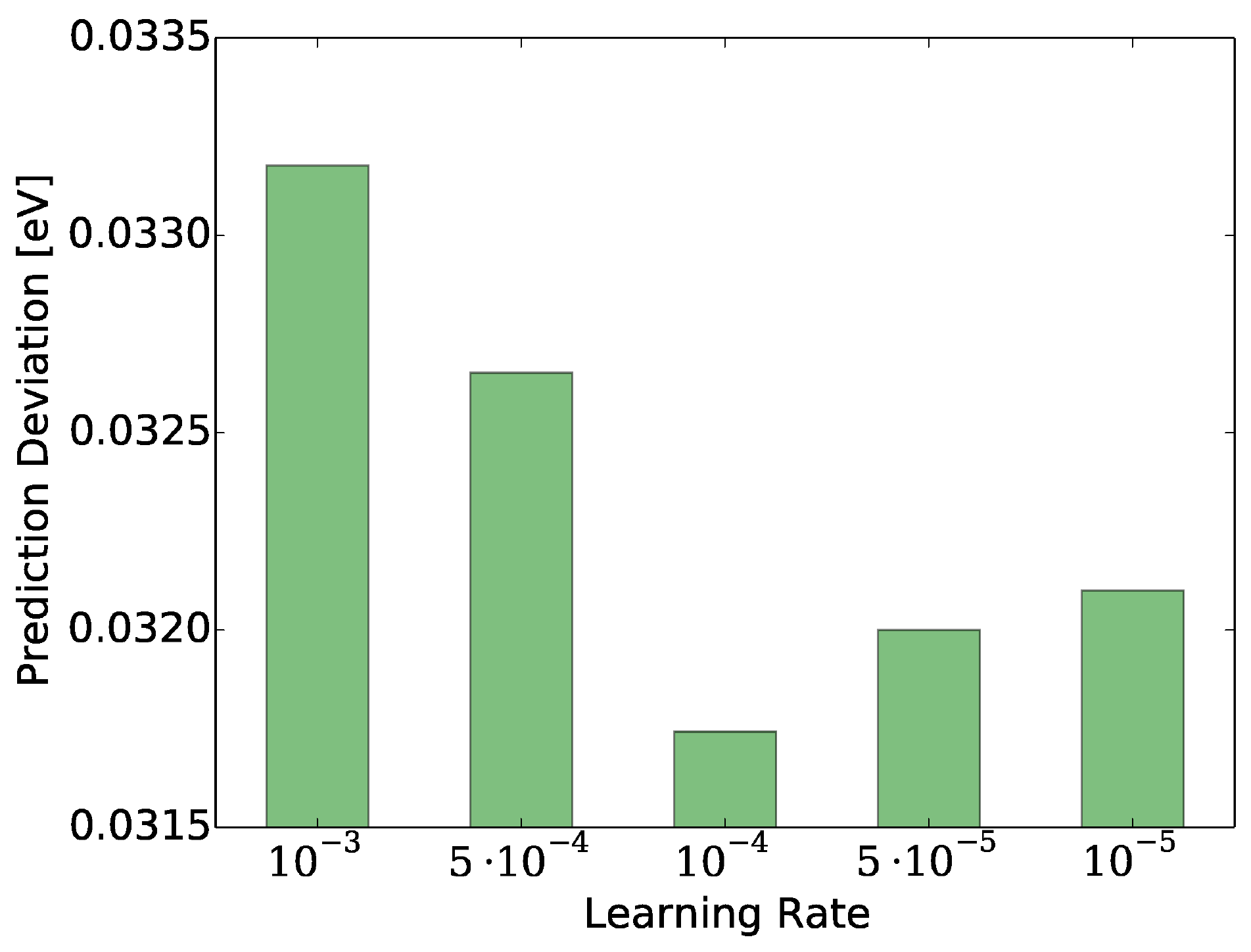}
\protect\caption{Average absolute deviations of predicted excited state energies from
TDDFT excited state energies for neural networks with different learning
rates. Neural networks were trained on $3000$ Coulomb matrices randomly
drawn from the $10000$ frame data set. A learning rate of $10^{-4}$
resulted in the lowest prediction error. However, prediction error
for other learning rates are less than $\unit[10]{\%}$ larger than
the minimal value. \label{fig:benchmarking_learningRate}}
\end{figure}

The average absolute deviations of predicted excited state energies
from TDDFT results for each neural network are depicted in Fig.~\ref{fig:benchmarking_learningRate}.
We see that deviations of single predicted excited state energies
from TDDFT excited state energies ranges from $\unit[31.8]{meV}$
for a learning rate of $10^{-4}$ to $\unit[33.2]{meV}$ for a learning
rate of $10^{-3}$. In any case, we found that the prediction error
increases as the learning rate deviates from $10^{-4}$. Hence, we
consider a learning rate of $10^{-4}$ to be optimal for the case
of excited state energies of BChls in the FMO complex. However, the
small changes of the prediction accuracy with different applied learning
rates indicates that the prediction accuracy of a neural network is
not very sensitive to the learning rate for this application.

\subsubsection*{Choice of number of neurons}

\begin{figure*}
\begin{centering}
\includegraphics[width=1\columnwidth]{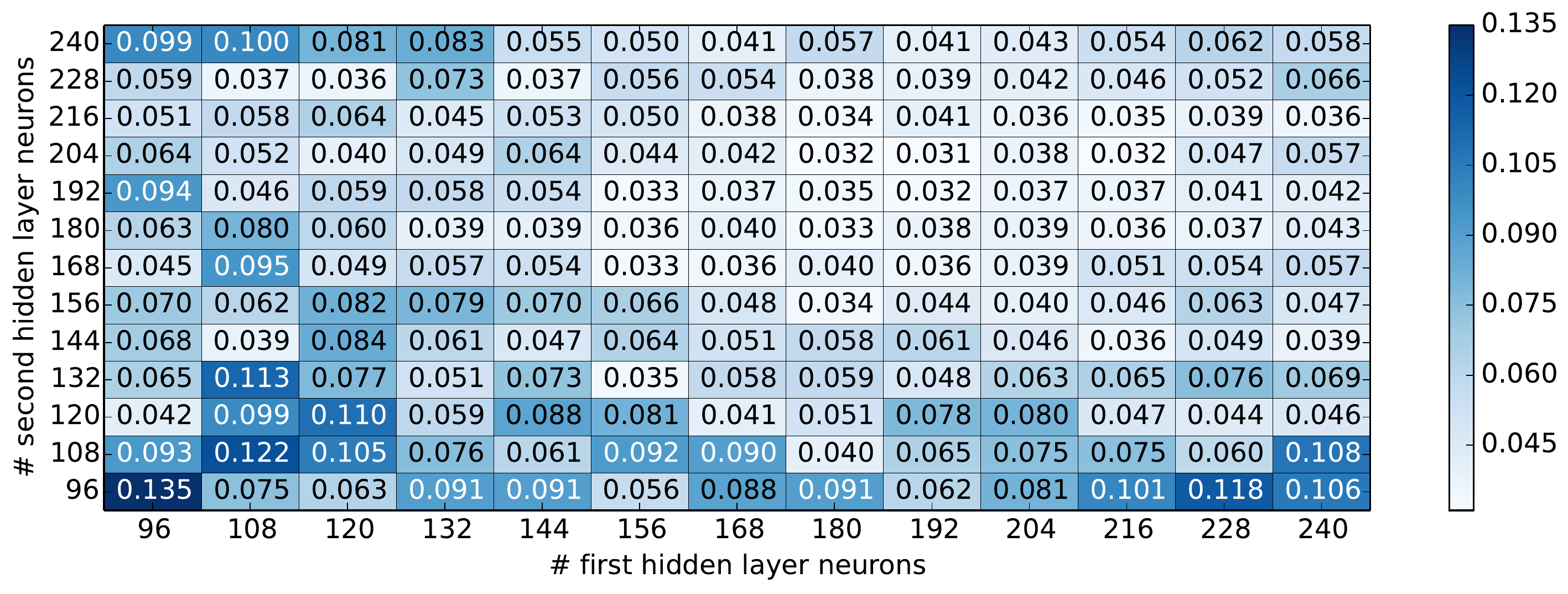}
\par\end{centering}

\protect\caption{Average absolute deviations of predicted excited state energies from
TDDFT calculated excited state energies for different numbers of neurons
in the first and second hidden layer. Deviations are reported in $\unit{eV}$.
Neural networks were trained on $3000$ Coulomb matrices randomly
drawn from the $10000$ frame data set. A hidden layer combination
of $204$ neurons in the first hidden layer and $192$ neurons in
the second hidden layer resulted in the smallest deviation of $\unit[31]{meV}$.
\label{fig:benchmarking_hiddenLayers}}
\end{figure*}

The influence of the number of neurons in the first and second hidden
layer on the prediction accuracy of a neural network was investigated
with several neural networks with a learning rate of $\unit[10]{^{-3}}.$
Neuron numbers in each of the two hidden layers were varied from $96$
to $240$ in steps of $12$. Each of the constructed neural networks
was trained on site 3 with $3000$ Coulomb matrices randomly selected
from the $10000$ frame data set. Results are illustrated in Fig.~\ref{fig:benchmarking_hiddenLayers}.
The neural network with $204$ neurons in the first hidden layer and
$192$ neurons in the second hidden layer showed the smallest average
absolute deviation of $\unit[31]{meV}$ between single predicted excited
state energies and TDDFT excited state energies. 

\subsubsection*{Training set size}

Aside from neural network hyperparameters we also investigated the
effect of the number of Coulomb matrices in the training set on the
prediction accuracy of the neural network. In general, the more data
is provided to the neural network during training, the more accurately
it can predict the targets. However, with the training set size the
computational time spent on neural network training also increases.
So we looked for a balance between the amount of data provided to
the neural network during training and the computational cost of the
neural network training. 
\begin{table}
\begin{centering}
\begin{tabular}{cc}
\toprule 
Training set size {[}$\unit{\#}$ frames{]} & Training time {[}$\unit{h}${]}\tabularnewline
\midrule
$500$ & $1.7\pm0.6$\tabularnewline
$1000$ & $3.8\pm1.4$\tabularnewline
$1500$ & $6.5\pm2.8$\tabularnewline
$2000$ & $13.6\pm6.5$\tabularnewline
$2500$ & $18.0\pm7.9$\tabularnewline
$3000$ & $20.3\pm5.5$\tabularnewline
$3500$ & $22.7\pm7.3$\tabularnewline
$4000$ & $23.9\pm5.0$\tabularnewline
$4500$ & $30.2\pm1.5$\tabularnewline
$5000$ & $31.2\pm4.8$\tabularnewline
\bottomrule
\end{tabular}
\par\end{centering}

\protect\caption{Training times for neural network training on four cores. Each neural
network was trained on Intel(R) Xeon(R) CPUs (X5650 @ $\unit[2.67]{GHz})$
with $\unit[4]{GB}$ of RAM. Training set sizes are reported in number
of frames randomly drawn from the $10000$ frames of the trajectory
of site 3. A total of $12$ neural networks was trained on each training
set size. Training times are reported with average and standard deviation
of all neural network training sessions with the particular training
set size. \label{tab:Training-times-for} }
\end{table}

Investigated training set sizes ranged from $500$ Coulomb matrices
to $5000$ in steps of $500$. In every case, Coulomb matrices of
site 3 were drawn randomly from the $10000$ frame data set. For each
training set size a total of $12$ neural networks was trained. Neural
networks had a learning rate of $10^{-4}$ with $204$ neurons in
the first hidden layer and $192$ neurons in the second hidden layer.
Recorded training times for different training set sizes are reported
in Tab.~\ref{tab:Training-times-for} as a $12$ neural network average.
Four cores of Intel(R) Xeon(R) CPUs (X5650 @ $\unit[2.67]{GHz})$
with $\unit[4]{GB}$ of RAM were used to train one neural network. 

We found that neural network training times significantly increase
when including more than $4000$ frames. Up to this training set size,
neural network training takes about one day, which we considered a
reasonable time for neural network training. Including $500$ more
frames to the training set increased the training time by about $\unit[6]{h}$
or $24$ core hours. As a balance of the amount of input data and
computational cost we therefore decided to use $4000$ frames in the
training set for neural network training.

\subsection{Spread of neural network predictions\label{sub:Spread-of-neural}}

\begin{figure}
\begin{centering}
\includegraphics[width=0.5\columnwidth]{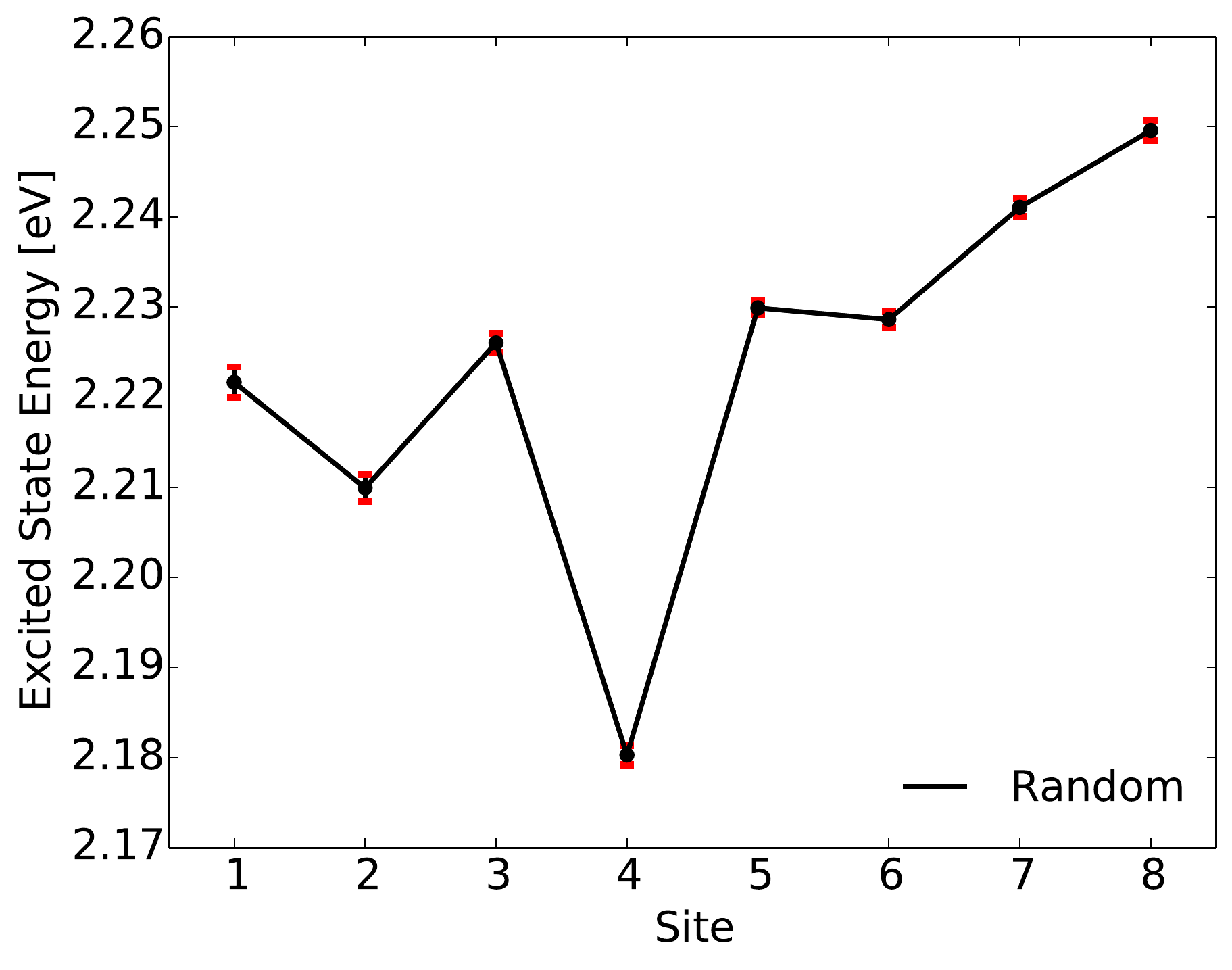}
\par\end{centering}

\protect\caption{Averages of neural network predicted excited state energy trajectory
averages. Error bars indicate the standard deviations of the neural
network predicted excited state energy trajectory distributions. A
total of $12$ neural networks was trained for each of the eight sites
on $4000$ Coulomb matrices randomly drawn from the $10000$ data
frames of the indicated site. \label{fig:Averages-of-neural}}
\end{figure}

We used a total of $12$ independent neural networks to predict excited
state energies for a particular BChl in the FMO complex. Predicted
trajectories of all $12$ neural networks were averaged to obtain
a more accurate predicted excited state energy trajectory. To justify
the usage of the average of predicted excited state energy trajectories
we report the spread of individual neural network predictions in Fig.~\ref{fig:Averages-of-neural}.
 Neural networks with optimal hyperparameters (see Sec.~\ref{sub:Neural-Network-Benchmarking})
were trained on $4000$ Coulomb matrices randomly drawn from the $10000$
frame data set for every respective site. After training, neural networks
were used to predict the entire excited state energy trajectory of
the site on which they were trained. Trajectory averages of predicted
excited state averages were calculated for every neural network prediction.
All predicted excited state energy trajectory averages were then ensemble
averaged over the $12$ neural networks trained on the particular
site. Results are reported in Fig.~\ref{fig:Averages-of-neural}
with the standard deviation of the neural network predicted excited
state energy trajectory averages as error bars.

\subsection{Excited state energy distributions\label{sub:Excited-state-energy}}

\begin{figure*}
\begin{centering}
\includegraphics[width=0.95\columnwidth]{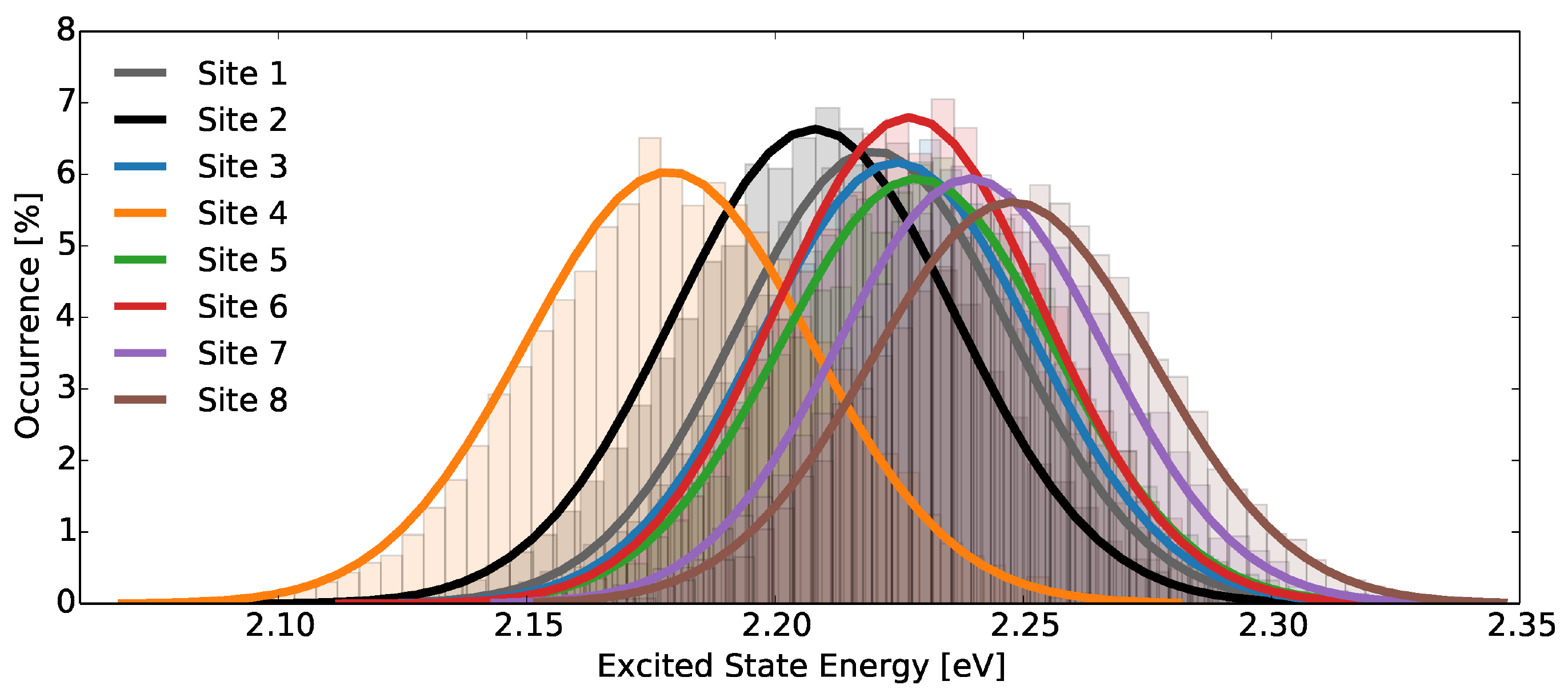}
\par\end{centering}

\protect\caption{Excited state energy distributions for all eight BChls in the FMO
monomer A. Excited state energies were obtained from TDDFT calculations
with the PBE0 functional and the 3-21G basis set. Distributions were
calculated from a total of $10000$ frames spanning $\unit[40]{ps}$
with a binning of $50$.\label{sup_fig:predictionErrors_full-2}}
\end{figure*}

Excited state energies were calculated for all BChl in the FMO at
every $\unit[4]{fs}$ of the $\unit[40]{ps}$ production run. Results
were obtained from TDDFT calculations with the PBE0 functional and
the 3-21G basis set using the Q-Chem quantum chemistry package. Distributions
of the obtained excited state energy trajectories are depicted in
Fig.~\ref{sup_fig:predictionErrors_full-2}.

\subsection{Spectral Densities and Exciton Dynamics\label{sub:Spectral-Densities}}

\subsubsection*{Spectral Densities for individual sites}

\begin{figure*}
\begin{centering}
\includegraphics[width=1\columnwidth]{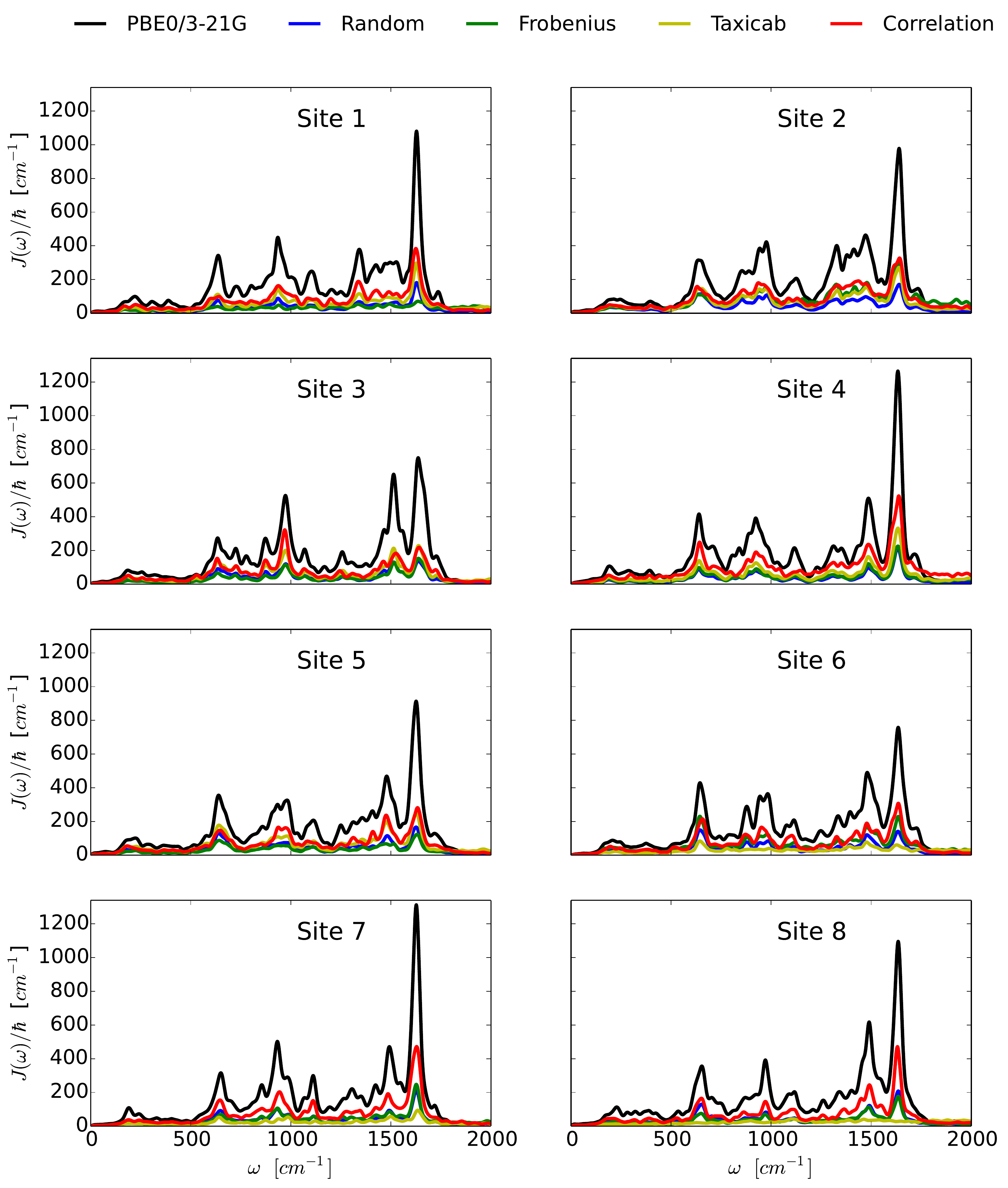}
\par\end{centering}

\protect\caption{Harmonic spectral densities for individual sites in the FMO complex.
The spectral densities were calculated from excited state energy trajectories
obtained from TDDFT calculations (PBE0/3-21G) and compared to spectral
densities from neural network predicted excited state energy trajectories.
Neural networks were trained on the BChl they predicted with the indicated
Coulomb matrix selection method. \label{fig:SD_supplementary}}
\end{figure*}

Spectral densities for individual BChls in the FMO complex were calculated
from TDDFT excited state energy trajectories and neural network predicted
excited state energy trajectories. Neural networks were trained on
Coulomb matrices selected with different selection methods from the
$10000$ frame trajectory. Harmonic spectral densities for individual
BChls are shown in Fig.~\ref{fig:SD_supplementary}.

For the average spectral density (see Fig.~\ref{fig:Spectral-density-average}
in Sec.~\ref{sub:Exciton-dynamics}) we observed that neural networks
trained on correlation clustered Coulomb matrices predicted spectral
densities significantly better than any other Coulomb matrix selection
method. However, the advantage of correlation clustering in terms
of spectral density prediction accuracy is less obvious for spectral
densities trained on individual BChls. 

Nevertheless, for each of the introduced Coulomb matrix selection
methods neural networks were able to predict the general shape of
the spectral density, although the area below the curves is significantly
smaller and correlation clustering identified more peaks than the
other selection methods. 
\begin{figure}
\begin{centering}
\textbf{A)}%
\begin{minipage}[t]{0.45\columnwidth}%
\begin{center}
\includegraphics[width=1\columnwidth]{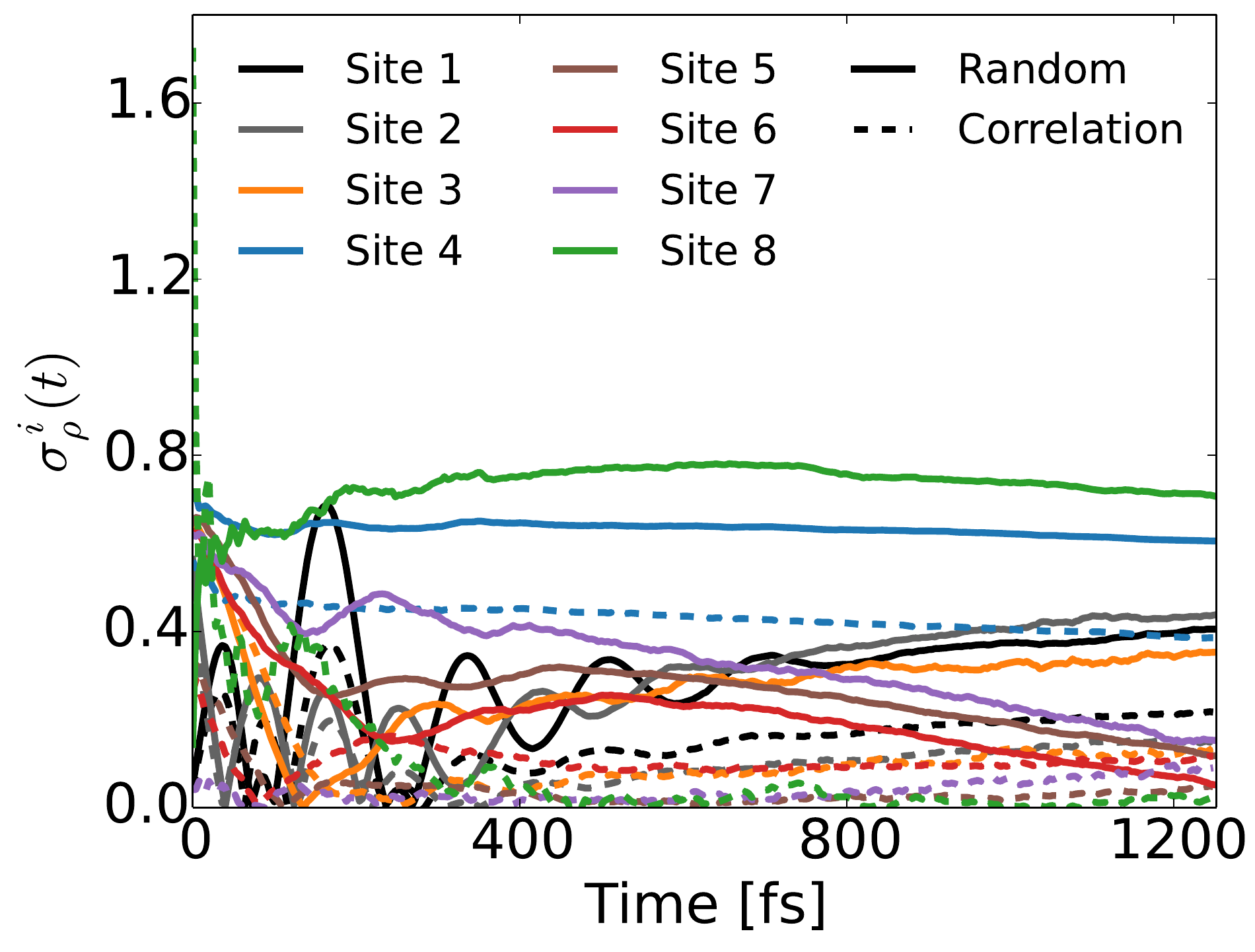}
\par\end{center}%
\end{minipage}\textbf{B)}%
\begin{minipage}[t]{0.45\columnwidth}%
\begin{center}
\includegraphics[width=1\columnwidth]{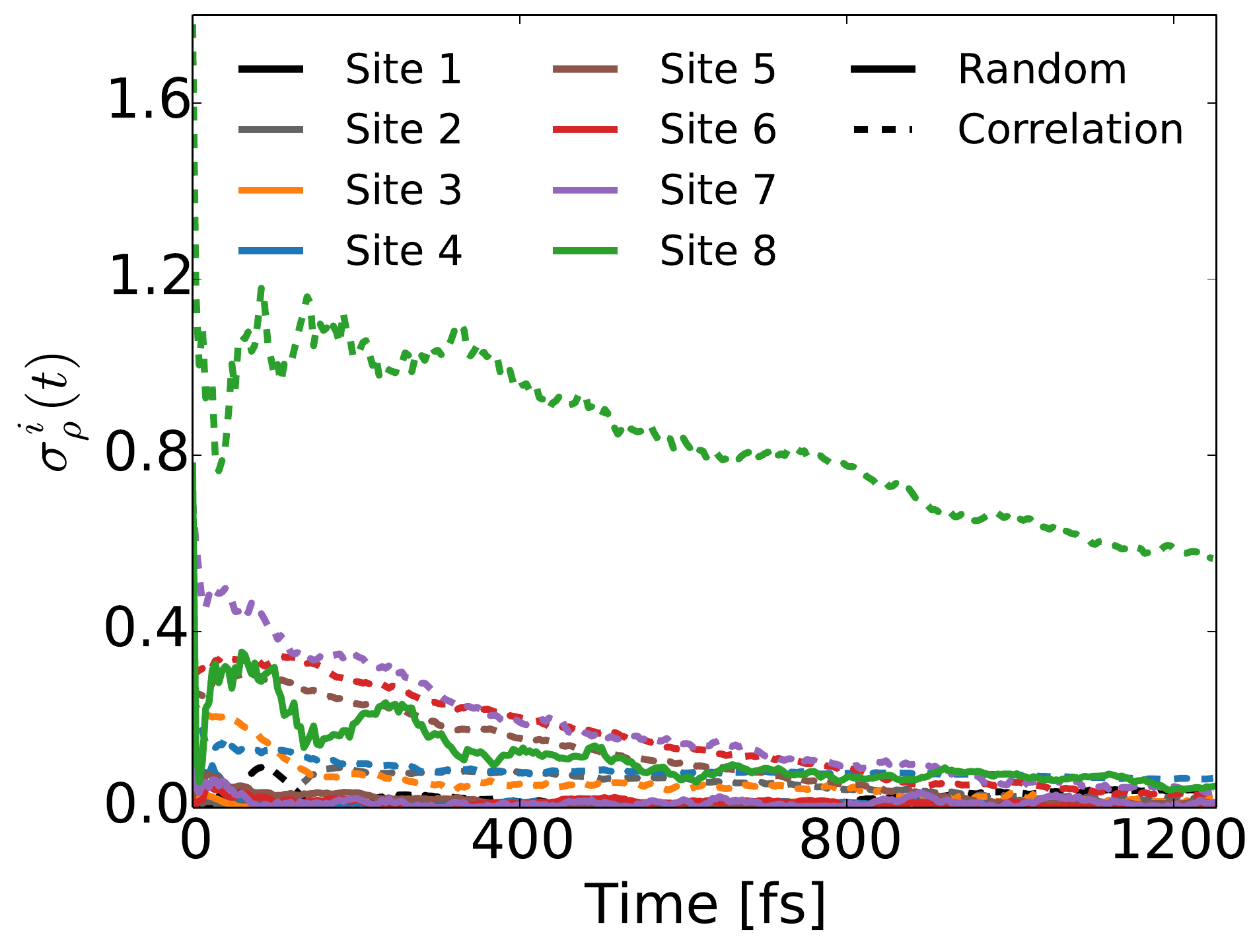}
\par\end{center}%
\end{minipage}
\par\end{centering}

\protect\caption{Deviation of TDDFT calculated exciton dynamics and neural network
predicted exciton dynamics over time. The deviation was calculated
as $\sigma_{\rho}^{i}(t)=|\rho_{ii}^{\text{TDDFT}}(t)-\rho_{ii}^{\text{NN}}(t)|/\rho_{ii}^{\text{TDDFT}}(t)$
with $i$ indicating the BChl. Panel A) shows the deviation for exciton
dynamics calculated with the Redfield method and neural network predicted
harmonic average spectral densities. In Panel B), the TDDFT average
spectral density was used for all calculations and neural networks
only predicted excited state energy trajectories. \label{fig:Deviation-of-TD-DFT}}
\end{figure}

\subsubsection*{Error estimation of predicted exciton dynamics}

The deviation of neural network predicted exciton dynamics and TDDFT
calculated exciton dynamics of the $i$-th BChl was quantified by
calculating $\sigma_{\rho}^{i}(t)=|\rho_{ii}^{\text{TDDFT}}(t)-\rho_{ii}^{\text{NN}}(t)|/\rho_{ii}^{\text{TDDFT}}(t)$.
Deviations were calculated for the exciton dynamics obtained with
the Redfield method as shown in Fig.~\ref{fig:Deviation-of-TD-DFT}.

We observe that for exciton dynamics calculations with neural network
predicted average spectral densities (see panel A in Fig.~\ref{fig:Deviation-of-TD-DFT})
the error of neural networks trained on randomly drawn Coulomb matrices
is significantly higher than the error of neural networks trained
on correlation clustered Coulomb matrices. This behavior can be observed
for all sites for times up to $\unit[1]{ps}.$ However, if we use
the TDDFT calculated average spectral densities for all simulations
instead of neural network predicted harmonic average spectral densities,
we observe a much smaller error for neural networks trained on randomly
drawn Coulomb matrices for all eight sites. This is due to the fact
that the error on the energies in slightly smaller with the random
sampling. 

\end{document}